\documentclass{article}

\usepackage{type1cm}        
\usepackage{makeidx}         
\usepackage{graphicx}        
\usepackage{multicol}        
\usepackage[bottom]{footmisc}
\usepackage{amsmath, amsfonts, amssymb, mathrsfs}

\usepackage{newtxtext}       %
\usepackage[varvw]{newtxmath}

\makeindex             

\begin{document}
\title{Nonminimal Higgs Inflation and Initial Conditions in Cosmology
\footnote{To the memory of V.A. Rubakov}}
\author{Andrei O. Barvinsky\footnote{Theory Department, Lebedev Physics Institute, Leninsky Prospect 53, Moscow 119991, Russia}\  and Alexander Yu. Kamenshchik\footnote{Dipartimento di Fisica e Astronomia ``Augusto Righi'',
Universit\`a di Bologna and INFN, via Irnerio 46, 40126 Bologna, Italy}}
\date{}

%
%
\maketitle


\abstract{We discuss applications of perturbative quantum gravity in the theory of very early quantum Universe and quantum cosmology. Consistency of the theoretical formalism for quantum effects of matter and correspondence with observational status of modern precision cosmology impose stringent bounds on and establish strong links with high energy particle phenomenology. Within this line of reasoning we study various aspects of one-loop approximation for the cosmological wave function, review Higgs inflation model intertwining the physics of electroweak sector of the Standard Model with the characteristics of the observable cosmic microwave background and, finally, consider the problem of quantum initial conditions for inflationary Universe. We formulate a cosmological quantum state in the form of the microcanonical density matrix -- a universal equipartition of eigenstates of the Wheeler-DeWitt equations. We demonstrate elimination of the inalienable infrared catastrophe of vanishing cosmological constant for the no-boundary quantum state of the Universe and derive initial conditions for inflation in the form of a special garland-type cosmological instanton -- the saddle point of quantum gravity path integral. Applied to the cosmological model of the Universe with a hidden sector of numerous conformally invariant higher spin fields, this setup suggests a solution to the problem of hierarchy between the Planck and the inflation energy scales and, thus, admits applicability of perturbative semiclassical expansion methods.}
\section{Introduction}
\label{sec:1}

Quantum Cosmology is an exiting field of modern theoretical physics. To begin with, it is an integral part of unified theory of elementary particles and fundamental interactions. As an application of quantum theory to the Universe as a whole, quantum cosmology is an inalienable part of quantization of gravity. Understanding the structure of physical interactions at the most fundamental microscopic scale of space and time, as is widely agreed now, is the prerogative of the advanced superstring theory motivated by intrinsic mathematical consistency and beauty along with the necessity of circumventing the known difficulties of quantizing Einstein general relativity. Microscopic realm of this problem seems to exclude quantum cosmology as a science whose field of application is the Universe in all its entirety. However, quantum origin of the early Universe is what brings macroscopic and microscopic phenomena together tightly intertwining their methods of description and, thus, transcending commonly accepted physics principle of separation of scales.

On the other hand, modern cosmology has overstepped the limits of the state of art and started profoundly describing various aspects of the evolution of the very early and present day Universe. As soon as it entered in early nineties the stage of precision cosmology it has become now observational science with a very solid experimental status. In particular, it deeply explained the phenomenology of hot Universe originating from the stage of a very fast quasi-exponential inflationary expansion. However, inflationary cosmology does not describe initial quantum state of the Universe -- primary goal of quantum cosmology. This is again the point at which complementary fields of physics embrace and call for unification of the presently observable large scale structure of the Universe with its quantum origin laying the imprint on this structure.

One can say that modern quantum cosmology began in 1967 when the famous Wheeler-DeWitt equation was put forward in \cite{DeWitt}. This equation arises as a result of application  to  gravity of the Dirac quantization procedure for systems with canonical constraints \cite{Dirac}. Briefly this equation, or it would be better to say the infinite systems of these equations -- four ones per spatial point, looks like
\begin{equation}
\hat{\cal H}|\Psi\rangle = 0,       \label{WDW}
\end{equation}
where $|\Psi\rangle$ is a quantum state of the Universe, which is sometimes called  a wave function of the Universe, while
$\hat{\cal H}$ is the set of Hamiltonian and momentum constraints of the canonical formalism of gravity theory. These constraints depend on spatial components of the spacetime metric and their conjugated momenta along with the full set of degrees of freedom of matter fields. This system is very complicated and its precise mathematically rigorous and explicit formulation is still missing because it involves unsolved issues of operator realization, operator ordering and consistency manageable only at the formal level.

But, as it often happens in physics, the absence of full rigourousness does not prevent from fruitful applications of imperfect mathematical tools by means of their extrapolation from those areas where they were well defined and productively exploited.\footnote{Sometimes it is even stated that the Wheeler-DeWitt equation is the most useless equation in physics. But this statement is certainly incorrect because, even if this equation is not directly used in concrete quantum gravitational applications, it still fundamentally underlies the results obtained by alternative methods, which are intrinsically equivalent to the Wheeler-DeWitt equation formalism. As a comparison one can point out at the Schroedinger equation which is hardly usable as a tool of high energy relativistic scattering but it fully underlies scattering phenomenology.} As such a tool we will use path integration method which gives a formal solution to the Wheeler-DeWitt equations \cite{Hartle-Hawk,Hawk,Bar}. Though this method does not fully resolve such conceptual problems of the formalism as the problem of time, probabilistic interpretation of the cosmological wavefunction \cite{Bar,Bar-Kam}, etc., it allows one to reach interesting predictions on the quantum origin of the Universe and its further evolution within a reliable scheme of covariant effective or renormalizable quantum field theory. Within semiclassical expansion this approach incorporates the concept of quantum mechanical tunneling or the ``birth or tunneling of the Universe from nothing'' as a natural interpretation of complex saddle points of the underlying path integral -- transition to the physical Lorentzian signature spacetime from the Euclidean manifold describing classically forbidden state of the gravitational field in imaginary time.

The implementation of this concept is the ``no-boundary'' or Hartle-Hawking prescription for the wave function of the Universe \cite{Hartle-Hawk,Hawk}. It is based on the direct application of the Euclidean quantum field theory to the full system of gravitational and matter fields in the Universe.
Semiclassically their quantum  state is described in terms of the imaginary
time, that is by means of the Euclidean spacetime, so that the corresponding amplitudes and probabilities are weighted by the exponentiated Euclidean gravitational action, $\exp(-S_E)$. The action is calculated on the gravitational instanton -- the saddle point of an underlying path
integral over Euclidean 4-geometries. This instanton gives rise to Lorentzian signature spacetime by analytic continuation across minimal hypersurfaces, this continuation being interpreted either as quantum tunneling or as the creation of the Universe from ``nothing''.  An immediate difficulty with this picture is the so-called infrared catastrophe of a small cosmological constant $\varLambda$ --- fundamental or effective (that is induced by matter fields and determining the size of the Universe). The simplest realization of the Hartle-Hawking wave function in minisuperspace approximation of the isotropic and homogeneous spacetime, which describes nucleation of the de Sitter Universe from the Euclidean 4-dimensional hemisphere, has the form
\begin{equation}
\Psi_{\rm HH} \sim \exp(-S_E) = \exp\left(\frac{3\pi}{2G\varLambda}\right).
\label{HH}
\end{equation}
This diverges for $\varLambda \rightarrow 0$ because of indefiniteness of the Euclidean gravitational action in Einstein general relativity. Such a result looks counterintuitive because it predicts as infinitely more probable the quantum birth of the Universe of infinitely big size.

The alternative tunneling state of the Universe \cite{Vil,Vil1,Linde,Rubakov,Star-Zeld} is based on the semiclassical solution of the minisuperspace version of the Wheeler-DeWitt equation -- that is the situation when the full set of Wheeler-DeWitt equations (\ref{WDW}) is truncated to a single differential equation for the wave function in the finite-dimensional Friedmann metric sector of the full space of 3-geometries. Setting boundary conditions for this equation appropriate for tunneling through classically forbidden region leads to the expression weighted by the same exponentiated Euclidean gravity action which is, however, taken with the opposite sign,
\begin{equation}
\Psi_{\rm T} \sim \exp(S_E) = \exp\left(-\frac{3\pi}{2G\varLambda}\right).
\label{T}
\end{equation}
This result looks less counterintuitive, but its essentially minisuperspace setup, which can hardly be formulated outside of the Friedmann metric context, gives a serious ground to doubt the fundamental nature of such a tunneling construction.

Apart from the foundational issues underlying the prescriptions (\ref{HH}) and (\ref{T}) they both suffer from the normalizability problem. If the effective cosmological ``constant'' is a composite function of some physical fields, like for example the effective potential of the inflaton field $\varLambda=V(\phi)$, both expressions are not suppressed in high energy domain $V(\phi)\to\infty$ and not normalizable quantum mechanically, $\int^\infty_{-\infty} d\phi\,|\,\Psi_{\rm HH,T}(\phi)|^2=\infty$ (in fact this is a counterpart to the infrared catastrophe -- non-integrable singularity of the wavefunction at $\varLambda=V(\phi)\to 0$ in the no-boundary case). So the formalism does not have intrinsic cutoff protecting semiclassical low-energy physics from the ultraviolet domain where semiclassical methods fail and nonperturbative methods are not yet available.

The further progress of quantum cosmology was associated with application of ideas and methods of the perturbative quantum gravity. Here, the word {\em perturbative} has several different meanings.  The first one is related to the extension beyond minisuperspace models and the inclusion of perturbations of the gravitational and matter fields on top of the Friedmann background \cite{Hal-Hawk, Laflamme}. The second aspect of the perturbative quantum gravity consists in the efforts to overcome the tree-level limit of the semiclassical expansion and to consider one-loop corrections to the wave function of the Universe \cite{Schleich,Louko,Esp-D'Eath,Esp-D'Eath1,BKKM, BKK,KM,KM1,KM2,EKMP,EKMP1,MP,Vas}. These works were focused on corrections to the no-boundary wave function of the Universe and, in particular, the relation between the covariant Schwinger-DeWitt formalism of local curvature expansion for quantum effective action \cite{Sch-DeWitt} and calculations on homogeneous spaces based on spectral summation and zeta-function regularization \cite{zeta-Hawk}.

In \cite{Bar-Kam1} the above two perturbative approaches were successfully combined by including the one-loop contribution of cosmological perturbations. Hartle-Hawking wave function of the Universe was computed in the one-loop approximation and rather generic effective action algorithm for the probability of inflation -- probability distribution of initial conditions for inflation -- was obtained. It was, in particular, shown that high energy normalisability of the cosmological wave function was determined by the anomalous scaling of the quantum theory on the cosmological gravitational instanton. The one-loop mechanism of generating local peaks of the inflation probability was proposed for the inflaton scalar field with a nonminimal gravitational coupling -- the model of inflation originally considered at the classical level in \cite{Vent,Spok,Salop,Fakir,Futa}. Thus, a direct link between quantum cosmology and particle physics content of the Universe was established.

Application of the obtained algorithm to tunneling wave function was considered in \cite{Bar-Kam2}. The corresponding quantum scale of inflation was found by observing a sharp probability peak in the distribution function of chaotic inflationary cosmologies driven by a scalar field with a large negative constant $\xi$ of nonminimal interaction
\cite{Bar-Kam-Mish,Bar-Kam3,Bar-Kam1,Bar-Kam2}. The attempt to justify this result within path integral representation of the tunneling wave function was undertaken in \cite{we-tun-Higgs} with regard to peculiarities of analytic continuation from the Euclidean geometry to the Lorentzian signature spacetime. This continuation was earlier studied in \cite{Bar-Kam-Mish} along with the derivation of effective equations of motion driving the inflation stage \cite{Bar-Kam3}.

Important issue in quantum cosmology (and generically in quantum theory) is an explanation of the fact that while our world has quantum origin, we observe it in a macroscopic context as a classical object -- quantum spreading does not destroy classical behavior of large bodies and let it be governed by principles of Laplace determinism. Generally agreed explanation of this fact, which otherwise would contradict simplest estimates of quantum mechanics, is the phenomenon of decoherence \cite{Zeh,Zurek}. Decoherence is the effective classicalization of the quantum world which arises due to the interaction of the observed physical variables with an unobservable cloud of degrees of freedom, which is usually called environment. The natural question which arises when one tries to explain the classicalization of a quantum Universe using the decoherence approach, is associated with the definition of such an environment. Indeed, in contrast to the usual description of quantum-mechanical experiment in a laboratory, there is no external environment, because the object of quantum cosmology is the whole Universe. Thus, we should treat a certain part of degrees of freedom as essential observables, while the rest of them should be considered as an environment with subsequent tracing them out within a formalism of reduced density matrix. It is natural to believe that inhomogeneous degrees of freedom play the role of environment while the macroscopic variables, such as a cosmological scale factor or initial value of the inflaton scalar field, should be treated as observables \cite{Kiefer}. Under these assumptions the reduced density matrix in cosmology was calculated in  \cite{Bar-Kam-Kief-Mish,Bar-Kam-Kief} by the method of \cite{Bar-Kam1,Bar-Kam2}.

Further progress in the theory of early Universe, at the overlap of particle physics, precision cosmology and quantum gravity is associated with the so-called Higgs inflation theory. Previously mentioned works on the nonminimally coupled to gravity scalar field, which plays the role of inflaton, explicitly or implicitly assumed that in cosmological context this field should belong to the scope of Grand Unification Theory (GUT). Rather challenging idea to identify this field with the Higgs boson (which was not yet discovered at the moment of the publication of this idea) was put forward in \cite{Bez-Shap}. The motivation for this transition from GUT to electroweak scale was an interesting hypothesis on nontrivial properties of renormalization group flow of the well-known Standard Model, extrapolating to the Planckian scale and somehow circumventing new physics in the GUT domain. However, a real interest in this model critically grew after the observation \cite{Bar-Kam-Star} that such an identification allows one to establish a connection between the observable cosmological parameters, such as the magnitude and spectral index of cosmic microwave background anisotropy, and the value of the Higgs boson mass -- the discovery of this first ever known scalar particle being eagerly expected at that time at LHC. The series of the following works \cite{Bez-Shap1,Bez-Shap2,Wil,Bar-Kam-Star1,Bar-Kam-Star2} using the renormalization group approach then gave more precise predictions for the relation between particle physics and cosmology of this Higgs inflation model. The further development of this model and its various versions can be found in the review paper \cite{Rubio}.

Success of Higgs inflation model, even though it was somewhat marred by controversy of the strong coupling scale \cite{Burgess} (later corrected in \cite{BMSS}), rather persuasively connects cosmology with the Standard Model of particle physics. However, Standard Model is often understood as an effective field theory, which at Planckian scale should follow from more fundamental paradigm of the superstring theory. It is well known that the variety of options open by string models -- landscape of string vacua -- is huge \cite{land,land1}. Thus, it sounds natural to find certain principles restricting this landscape, and these principles might stem not only from the internal logic of string theory but, in particular, might follow from quantum cosmology.

On the other hand, Higgs inflation does not explain the origin of initial conditions for inflationary scenario. Obviously, this should be the domain of energy scales where semiclassical physics of quantum gravity and quantum theory of Standard particle physics model both meet with the frontier of essentially nonperturbative quantum gravity. So, in the absence of nonperturbative methods, the question arises whether we still can rationally describe quantum origin of the Universe and whether the no-boundary or tunneling prescriptions for its quantum state can survive extension to relevant energy scales. The answer, of course, depends on what is the energy scale of this phenomenon -- if it is sufficiently below the Planck scale (and the interpretation of the current cosmological data from Planck satellite \cite{Planck} is such that it is indeed four, five or even more orders of magnitude below Planckian $10^{19}$ GeV) than we can hope on the success of our theoretical description. But the energy scale of the quantum birth of the Universe should not be input by hands -- to be a reliable element of the theoretical scheme it should follow from the physical model of our Universe and its particle physics contents.

One step in this direction, which might resolve the above issue and simultaneously supply string theory with possible landscape selection criteria, was undertaken in \cite{Bar-Kam-land,Bar-Kam-land1,Bar-Land}. The main guiding rule in the implementation of this step was due to Occam razor principle to avoid unnecessary and redundant assumptions on the choice of distinguished quantum states like the no-boundary or tunneling ones and replace this choice by a universal equipartition -- the microcanonical density matrix of the Universe \cite{Bar-Land}. Initially interpreted in \cite{Bar-Kam-land,Bar-Kam-land1} as an extension of the old idea of quasi-thermal Euclidean quantum gravity density matrix \cite{Page}, this suggestion was understood in \cite{Bar-Land} as equipartition in the space of all wavefunctions satisfying the system of Wheeler-DeWitt equations. This is the analogue of a conventional microcanonical density matrix for which, however, even the notion of physical time fundamentally arises as an operator ordering parameter. Application of this idea to the cosmological model with many species of conformally invariant quantum fields leads to a number of remarkable conclusions. In particular, it eliminates the inalienable infrared catastrophe of the no-boundary state, restricts from above the cosmology energy scale -- the main ground for a possible landscape selection rule, determines the CMB spectrum and energy scale of inflation \cite{Bar-Kam-Nes,Bar-Kam-Nes1} depending on the tower of conformal particles in the model and, thus, establishes the hierarchy between this scale and the effective Planckian cutoff below which we can safely apply semiclassical perturbation theory \cite{CHS}. Other interesting properties relating this model to other gravitational models can be found in \cite{Bar-Kam-Def,Bar-Kam-Def1}.

In this paper we would like to give a comprehensive review of the most interesting ideas and the methods briefly reviewed above. The structure of paper is the following: the second section presents the wave function of the Universe in the one-loop approximation; the third section gives details of the nonmininal Higgs inflation model; in the fourth section the theory of the cosmological density matrix is presented; the last section contains concluding remarks.

\section{Wave function of the Universe in the one-loop approximation}

The general theory of the wavefunction in quantum cosmology, as a solution of the system of Wheeler-DeWitt equations, and the way how this theory can be embedded in the canonical quantization formalism of constrained dynamical systems was essentially advanced in the series of papers \cite{gen-sem,Bar,ordering1,ordering2,path-int1,path-int2}, including detailed one-loop approximation for quantum Dirac constraints \cite{gen-sem}, operator ordering problem and the problem of physical inner product \cite{ordering1,ordering2}, realization of the path integral method for such systems \cite{path-int1,path-int2}, unitarity in quantum cosmology \cite{Bar}, etc. These works, though they illuminate many hard issues in quantum cosmology, still remain at the formal level of the state of art and do not lead to concrete physical results. One of the reasons is that these works disregard infinite dimensional nature of gravitational configuration space and rely on manifestly noncovariant canonical formalism incapable of handling the issue of ultraviolet renormalization, quantum anomalies and so on. So, in this section instead of considering formal issues of quantum cosmology we focus on the one-loop approximation for the cosmological wavefunction, in which these issues of quantum state normalization, anomalies, etc. arise in full height.

\subsection{Normalizability of cosmological wavefunctions}
The no-boundary and tunneling cosmological wavefunctions with the contribution of linearized field fluctuations on top of the Friedmann homogeneous metric background read as infinite products of Gaussian (quasi-vacuum) states of harmonic oscillators with a time-dependent frequency parameter,
\begin{equation}
\Psi(t\,|\,\varphi,f)=\frac{1}{\sqrt{v^*_{\varphi}(t)}}\exp\Big(\mp I(\varphi)/2+iS(t,\varphi)\Big)\times\prod_n\psi_n(t,\varphi\,|\,f_n),
\label{dec-cosm}
\end{equation}
\begin{equation}
\psi_n(t,\varphi\,|\,f_n)=\frac{1}{\sqrt{v_n^*(t)}}
\exp\left(-\frac12\Omega_n(t)f_n^2\right),\quad
\Omega_n(t)=-a^k(t)\frac{\dot{v}_n^*(t)}{v_n^*(t)}.
\label{dec-cosm2}
\end{equation}
Here, the sign minus or plus in front of the Euclidean
action $I(\varphi)$  in the exponential of \eqref{dec-cosm} corresponds to
the no-boundary  and to the tunneling  wave
functions of the Universe, respectively, $f_n$ describe
amplitudes of inhomogeneous modes, while $v_n$
correspond to positive-frequency solutions of linearized second-order differential equations for these modes. The power $k$ of the cosmological scale factor $a(t)$ in the expression for the function $\Omega_n$, depends on the spin $s$ of the field under consideration and on its parametrization. For the ``standard'' parametrization $k =3 - 2s$. Inclusion of inhomogeneous modes into the wave function of the Universe was first considered
in \cite{Hal-Hawk, Laflamme}.
Principal (and rather nontrivial) achievement of \cite{Bar-Kam1} was that the diagonal of the reduced density matrix
corresponding to the wave function \eqref{dec-cosm}
\begin{equation}
\rho(t\,|\,\varphi)\equiv\rho(t\,|\,\varphi,\varphi) = \int\prod_n df_n|\,\Psi(t\,|\,\varphi,f)\,|^2
\label{dec-cosm3}
\end{equation}
can be represented by the one-loop effective action of the full set of inhomogeneous field modes $\varGamma_{\rm 1-loop}$ which admits covariant regularization and renormalization of its UV divergences,
\begin{equation}
\rho(t\,|\,\varphi) = \frac{\sqrt{\Delta_{\varphi}}}{|\,v_{\varphi}(t)\,|}\exp\Big(\!\mp I(\varphi)-\varGamma_{\rm 1-loop}(\varphi)\Big).
\label{dec-cosm4}
\end{equation}
Here
\begin{equation}
\Delta_{\varphi}\equiv ia^k(v_{\varphi}^*\dot{v}_{\varphi}-\dot{v}_{\varphi}^*v_{\varphi}).
\label{dec-cosm5}
\end{equation}
is the Wronskian of the basis functions of the wave operator of the field $\varphi$ and $\varGamma_{\rm 1-loop}(\varphi)$  is this one-loop effective action calculated on the DeSitter instanton of the radius $1/H(\varphi)$, where $H(\varphi)$ is the effective Hubble parameter as a function of the inflaton field. When $H(\varphi) \rightarrow \infty$, the covariantly regularized and renormalized effective action of the infinite set of inhomogeneous quantum modes $\varGamma_{\rm 1-loop}(\varphi)$ has a scaling behavior
\begin{equation}
\varGamma_{\rm 1-loop}(\varphi) = Z\ln\frac{H(\varphi)}{\mu},
\label{dec-cosm6}
\end{equation}
where $Z$ is the anomalous scaling of the theory, $\mu$
is a renormalization scale. Then the condition of normalizability of the
wave function of the Universe in UV domain reduces to the requirement that the parameter $Z$ should be bounded from below by a positive number depending on the sector of the homogeneous field mode $\varphi$. In this particular case this bound is
\begin{equation}
Z > 1,
\label{dec-cosm7}
\end{equation}
and this condition provides us with the selection
criterium for particle physics models \cite{Bar-Kam1}.
The work \cite{Bar-Kam1} was limited to the case of the Hartle-Hawking (no-boundary) wave function, and it was mentioned there that the extension to the model of chaotic inflation driven by a scalar field nonminimally coupled to gravity can bring serious advantages in comparison with the inflation driven by the minimally coupled scalar field.

Comparison of one-loop tunneling wave function of the Universe with the no-boundary wave function  for a nonminimally coupled
scalar field \cite{Bar-Kam2} begins with the classical Lagrangian in the Jordan frame of fields
\begin{eqnarray}
L(g_{\mu\nu},\varphi) = g^{1/2}\left[\frac{m_P^2}{16\pi}R-\frac12\xi R\varphi^2-\frac12(\nabla\varphi)^2
-\frac12m^2\varphi^2-\frac{\lambda}{4}\varphi^4\right].
\label{nonmin}
\end{eqnarray}

For a negative nonminimal coupling constant $\xi = -|\,\xi\,|$  this model can generate chaotic inflationary scenario with the inflaton potential in the Einstein frame
\begin{equation}
U(\phi)|_{\phi = \phi(\varphi)}=\frac{m^2\varphi^2/2
+\lambda\varphi^4/4}{(1+8\pi|\xi|\varphi^2/m_p^2)^2},
 \label{nonmin1}
 \end{equation}
and the one-loop approximated probability distributions in the no-boundary and tunneling states acquire the following form
\begin{equation}
\rho_{NB, T}(\varphi)=
\exp\left[\pm\frac{3m_P^4}{8U(\phi(\varphi))}\right]
\varphi^{-Z-2}.
\label{nonmin2}
\end{equation}
At least naively, this makes the both no-boundary and tunnelling
wavefunctions normalizable at over-Planckian scales
provided the parameter $Z$ satisfies the inequality $Z >
- 1$ serving as a selection criterion for consistent particle
physics models with a justifiable semiclassical loop
expansion.

Although the expression (\ref{nonmin2}) is strictly valid only in the limit $\varphi \rightarrow \infty$, it can be used for a qualitatively good description at intermediate energy
scales. In this domain the distribution (\eqref{nonmin2}  can generate
the inflation probability peak at $\varphi=\varphi_I$ with the dispersion $\sigma$,
\begin{eqnarray}
&&\varphi_I^2 = \frac{2|I_1|}{Z+2},\quad \sigma^2 = \frac{|I_1|}{(Z+2)^2},\nonumber \\
&&I_1 = -24\pi\frac{|\xi|}{\lambda}(1+\delta)m_P^2,\quad \delta = -8\pi\frac{|\xi|m^2}{m_P^2}.
\label{nonmin3}
\end{eqnarray}
Here, $I_1$ is a second coefficient of expansion of the
Euclidean action in inverse powers of $\varphi$:
\begin{equation}
I(\varphi) = -\frac{3m_P^4}{U(\phi(\varphi))}=I_0+\frac{I_1}{\varphi^2} + O(1/\varphi^4).
\label{nonmin4}
\end{equation}
For
the no-boundary and tunnelling states this peak exists
in complimentary ranges of the parameter $\delta$. For the no-boundary state it can be realized only
for $\delta < -1 (I_1 > 0)$ and, thus, corresponds to
the eternal inflation with the field $\varphi$ on the negative
slope of the inflaton potential \eqref{nonmin1} growing
from its starting value $\varphi_I$. For a tunnelling
proposal this peak takes place for $\delta > -1$ and
generates the finite duration of the inflationary
stage.

We can ask ourselves how the anomalous scaling $Z$ behaves in the theories with strong negative nonminimal coupling and at large values of the scalar field $\varphi$. It was well known  \cite{Allen,Frad-Tseyt} that the expression for $Z$ includes the quartic contributions in terms of effective masses of particles present in the model. For a generic theory on a spherical de Sitter background
\begin{equation}
Z = \frac{1}{12H^4}\Big(\sum_{\chi}m_{\chi}^4+4\sum_{A}m_A^4
-4\sum_{\psi}m_{\psi}^4\Big),
\label{nonmin5}
\end{equation}
where $H$ is the inverse de Sitter radius (or the Hubble parameter) and the summation goes
over all  scalars $\chi$, vector gauge bosons $A$ and
Dirac spinors $\psi$. Their effective masses for large $\varphi$ are
dominated by the contributions $m^2_{\chi} = \lambda_{\chi}\phi^2/2, m^2_A =
g_{A}\varphi^2$ and $m_{\psi}^2 = f_{\psi}^2\varphi^2$ induced via the Higgs mechanism
from their interaction Lagrangian with the inflaton field.

Thus, in view of the relation $\varphi^2/H^2 = 12|\xi|/\lambda$, we
get the leading contribution of large $|\xi|$ to the total
anomalous scaling of the theory:
\begin{equation}
Z = 6\frac{\xi^2}{\lambda}{\bf A}+O(|\xi|),
\label{nonmin6}
\end{equation}
where
\begin{equation}
{\bf A} = \frac{1}{2\lambda}\left(\sum_{\chi}\lambda_{\chi}^2
+16\sum_{A}g_A^4-16\sum_{\psi}g_{\psi}^4\right),
\label{nonmin7}
\end{equation}
which contains the same large dimensionless ratio
$\frac{\xi^2}{\lambda}$ and the universal quantity $\bf A$ determined
by a particle physics model.

Thus, the consideration of the wave function of the Universe in the one-loop approximation establishes interesting link between the cosmology and particle physics, it permits to make both the no-boundary and tunneling wave functions of the Universe normalizable and for the case of the tunneling wave function it produces the probability distribution peak predicting initial conditions for inflation.

\subsection{Decoherence in quantum cosmology}
The perturbative technique sketched above also allows one to study the off-diagonal elements of the reduced density matrix of the universe obtained from the no-boundary or tunneling wave functions \cite{Bar-Kam-Kief-Mish,Bar-Kam-Kief} by integrating out the inhomogeneous fields. This is interesting because it gives the information about the decoherence of the Universe which is responsible for its classicalization. As we have already mentioned in the Introduction, the role of environment in the decoherence approach to cosmology is played by the part of the degrees of freedom, which are, generally, higher-order harmonics representing the inhomogeneous cosmological perturbations. Tracing them out we obtain the reduced density matrix. Information about the decoherence behaviour
of the system is contained in the off-diagonal elements
of its density matrix. In our case they read
\begin{equation}
\rho(t\,|\,\varphi,\varphi')=
\left(\frac{\Delta_{\varphi}
\Delta_{\varphi'}}{v_{\varphi}v_{\varphi'}^*}\right)^{1/4}
\exp\left(-\tfrac12\varGamma-\tfrac12\varGamma'
+i(S-S')\right)D(t\,|\,\varphi,\varphi').
\label{dec-cosm8}
\end{equation}
Here $\varGamma=\varGamma(\varphi)$, $\varGamma'=\varGamma(\varphi')$, $S$ and $S'$ are Lorentzian counterparts to these Euclidean actions corresponding to the overbarrier evolution from the Euclidean-Lorentzian transition points to the points $\varphi$ and $\varphi'$ and $D(t|\varphi,\varphi')$ is the so called decoherence factor defined by the formula
\begin{equation}
D(t\,|\,\varphi,\varphi')=\prod_n\left(\frac{4{\rm Re}\,
\Omega_n {\rm Re}\,\Omega_n'^*}{(\Omega_n+\Omega_n'^*)^2}\right)^{\frac14}
\left(\frac{v_n\,v_n'^*}{v_n^*\,v_n'}\right)^{\frac14}.
\label{dec-cosm9}
\end{equation}

How to cope with ultraviolet divergences appearing
in the infinite product of this type? This question was already discussed in \cite{Paz, Kiefer1, Okamura}. In \cite{Bar-Kam-Kief-Mish,Bar-Kam-Kief} we  used the dimensional regularization \cite{dimen}. As usual, the main effect of a dimensional regularization consists in changing the number of degrees of freedom involved in summation. For example, for a scalar field, the
degeneracy number of harmonics in spacetime
of dimensionality $d$ changes from the well-known value (see e.g. \cite{Khalat}) ${\rm dim}(n, 4) = n^2$, to
\begin{equation}
{\rm dim}(n, d) = \frac{(2n+d-4)\varGamma(n+d-3)}{\varGamma(n)\varGamma(d-1)}.
\label{dec-cosm11}
\end{equation}
Making analytical continuation and discarding
the poles $1/(d-4)$ one has finite values
for $D(t|\varphi,\varphi')$. However, for scalar, photon and
graviton fields one gets because of oversubtraction of UV infinities a pathological behaviour:
\begin{equation}
|\,D(t\,|\,\varphi,\varphi')\,| \rightarrow \infty,\ {\rm at}\ |\varphi-\varphi'| \rightarrow \infty.
\label{dec-cosm12}
\end{equation}
For example, for a massive scalar field $|D(t|\varphi,\varphi')|\approx \exp\big(\tfrac{7}{64}m^3\bar{a}(a-a')^2\big)$, where
\begin{eqnarray}
&&a=\frac{1}{H(\varphi)}\cosh H(\varphi)t,\quad
a'=\frac{1}{H(\varphi')}\cosh H(\varphi')t,\quad
\bar{a}=\frac{a+a'}{2}.
\label{dec-cosm14}
\end{eqnarray}
Such a form of a decoherence factor not only does
not correspond to decoherence, but also renders
the density matrix ill defined, breaking the condition
${\rm Tr}(\rho^2) \leq 1$.
However, there is a remedy -- using
the reparametrization of a bosonic scalar field
\begin{equation}
f \rightarrow \tilde{f} = a^{\mu}f,\quad v_{n} \rightarrow \tilde{v}_n=a^{\mu}v_n,
\label{dec-cosm15}
\end{equation}
one can get the new set of frequency functions
\begin{equation}
\Omega_n(t)=-ia^{3-2\mu}(t)\frac{\dot{\tilde{v}}_n^*(t)}{\tilde{v}_n^*(t)}.
\label{dec-cosm16}
\end{equation}
In such a way one can suppress ultraviolet divergences.
For the so-called conformal parametrization,
$\mu = 1$, for the massive scalar field one has
\begin{equation}
|\,\tilde{D}(t\,|\,\varphi,\varphi')\,|
=\exp\left(-\frac{m^3\pi\bar{a}(a-a')^2}{64}\,\right),
\label{dec-cosm17}
\end{equation}

For the case of fermions this method does not work \cite{Bar-Kam-Kief}. The wave function of the Universe filled with fermions has the form \cite{D'Eath-Hal, Kiefer-ferm, Bar-Kam-Kief}
\begin{eqnarray}
&&\Psi(t,\varphi|x,y) = \Psi_0(t,\varphi)\prod_n\psi_n(t|x_n,y_n),
\label{dec-cosm18}\\
&&\psi_n(t|x_n,y_n)=v_n-\frac{i\,\dot{v}_n+\nu\, v_n}{m}x_ny_n, \label{dec-cosm19}
\end{eqnarray}
where $x$ and $y$  are Grassmann variables and partial wave functions read in terms of relevant basis functions $v_n$ satisfying the second-order equation
\begin{equation}
\ddot{v}_n+(-i\dot{\nu}+m^2+\nu^2)v_n=0,\quad \nu = \frac{n+\frac12}{a}.
\label{dec-cosm20}
\end{equation}
The corresponding result reads in terms of the exponentiated divergent sum \cite{Kiefer-ferm}
\begin{equation}
|D(a,\varphi|a',\varphi')|=\exp\left(-\frac{m^2(a-a')^2}{8}
\sum_{n=1}\frac{n(n+1)}{\big(n+\tfrac12\big)^2}\right),
\label{dec-cosm21}
\end{equation}
whose dimensional regularization can be done using
the fact that for spinors in $d$-dimensional spacetime the degeneracy of the spectrum eigenvalues is equal to
\begin{equation}
{\rm dim}(n,d)=\frac{\varGamma(n+2^{(d-2)})
\varGamma(n+2^{(d-2)/2}-1)}{[\varGamma(2^{(d-2)/2})]^2
\varGamma(n+1)\varGamma(n)}.
\label{dec-cosm22}
\end{equation}
Still the renormalization procedure fails, like in the boson case, because it gives non-integrable kernel of the reduced density matrix with growing off-diagonal elements,
\begin{equation}
|\,D(a,\varphi|a',\varphi')\,|
=\exp\left(-m^2(a-a')^2\,I\right),
\label{dec-cosm23}
\end{equation}
having a negative finite constant $I<0$. Moreover, one cannot
use the conformal reparametrization in this case
because standard fermion variables are already
presented in the conformal parametrization.

However, there is another way to circumvent this problem.
One can perform a non-local Bogoliubov transformation
mixing Grassmann variables $x$ and $y$.
This transformation modifies the effective mass of
fermions in equation (\ref{dec-cosm20}) for their basis functions.
Choosing it in a certain way one can
suppress ultraviolet divergences. The reasonable
idea is to fix this transformation by the requirement that the decoherence is absent in a static spacetime. Then one gets a satisfactory result (\ref{dec-cosm23}) with a positive finite $I$.

All these examples imply strong dependence on parameterization of quantum variables and cast certain doubt on physical significance of the results. However, these examples show that consistency of the reduced density matrix might determine the very definition of the environment in a quantum cosmological model and serve as a criterion of the choice of observable (pointer) basis in field-theoretical systems -- a serious issue of the decoherence theory.

\section{Higgs boson, renormalization group, and naturalness in
cosmology}

\subsection{Higgs inflation model}
In the works considered in the preceding section the non-minimally coupled inflaton scalar field was associated with one of the scalars arising in Grand Unification Theories (GUT). Pioneering idea that the nonminimally coupled inflaton is nothing but the Higgs field of electroweak sector of the Standard Model was put forward in \cite{Bez-Shap}. The Lagrangian of this model in the graviton-inflaton sector reads,
\begin{eqnarray}
    &&L(g_{\mu\nu},\Phi)=
    \frac12\left(M_{\rm P}^2+\xi|\Phi|^2\right)R
    -\frac{1}{2}|\nabla\Phi|^{2}
    -V(|\Phi|),                     \label{inf-grav}\\
    &&V(|\Phi|)=
    \frac{\lambda}{4}(|\Phi|^2-v^2)^2,\,\,\,\,
    |\Phi|^2=\Phi^\dag\Phi,
    \label{EPJC}
    \end{eqnarray}
where $\Phi$ is a scalar field multiplet, whose expectation value plays
the role of an inflaton and which has a strong non-minimal curvature
coupling with $\xi\gg 1$. Here, $M_{\rm P}=m_{\rm P}/\sqrt{8\pi}\approx
2.4\times 10^{18}$ GeV is a
reduced Planck mass, $\lambda$ is a quartic self-coupling of
$\Phi$, and $v$ is a symmetry breaking scale.

It was advocated in \cite{Bez-Shap} that for the case when the scalar field is the Higgs field
the corresponding CMB data are consistent with the WMAP observations in
the tree-level approximation of the theory. The further history of
this non-minmally coupled Higgs inflation model was as follows.
The methods of \cite{Bar-Kam2,Bar-Kam3,Bar-Nes} were used to extend
the predictions in this model to the one-loop level \cite{Bar-Kam-Star}.  This
has led immediately to the lower bound on the Higgs mass $M_{\rm H}\approx 230$
GeV, originating from the observational restrictions on the CMB
spectral index \cite{Bar-Kam-Star}. However, this conclusion did not take into
account $O(1)$ effects of the renormalization group (RG) running,
which qualitatively change the situation. This was nearly
simultaneously observed in \cite{Wil} and in \cite{Bez-Shap1}, where the RG
improvement of the one-loop results of \cite{Bar-Kam-Star} has decreased the
lower bound on the Higgs mass to about $135$ GeV.

Quantitatively, this result was confirmed in \cite{Bar-Kam-Star1},
where we suggested the RG improvement of our one-loop results in
\cite{Bar-Kam-Star} and found a range of the Higgs mass that is compatible with the CMB. Both the lower and upper boundary of this range
were determined by the lower WMAP bound on the CMB
spectral index -- its value accepted at that time to be $n_s\approx 0.94$. The predictions of this model have also been extended to the two-loop approximation \cite{Wil,Bez-Shap2}, which has led to a reduction of
the lower bound on the Higgs mass range by about 10
GeV, which nearly coincides with the observed by LHC value of 125 GeV.

Simultaneously with the papers advocating Higgs inflation, including
in particular its supersymmetric extension \cite{supersym,KS12}, there
arose a number of objections to this model. Apart from the strong
assumption that no ``new physics" is present between electroweak (EW) and
inflation scales, it was criticized that the predictions for Higgs
inflation rely on perturbation
theory, which is only valid below the strong coupling scale. The reason for this criticism is that, for flat
space perturbation theory with a vanishing Higgs field background, this
scale turns out to be $M_{\rm P}/\xi$, which is much lower than the
inflation scale $M_{\rm P}/\sqrt\xi$
\cite{BurgLeeTrott1,Barbon,BMSS}, rendering the application of
perturbation theory questionable. Moreover,
the multi-component nature of the Higgs
field leads to the impossibility of canonically normalizing all
its components in the Einstein frame \cite{BurgLeeTrott1,Hertzberg,Kaiser} -- the
parametrization heavily employed in \cite{Wil,Bez-Shap2}.

In the approach of asymptotically safe gravity \cite{Weinberg,asympsafegrav,asympsafegrav1,asympsafegrav2},
one has also started to discuss the Higgs inflation model. There,
however, one has not incorporated the model (\ref{inf-grav}) with large $\xi$, but has exploited a rather miraculous numerical observation -- a certain relation between the EW instability in the SM and
the Planck scale \cite{asympsafe,BezShap4}.\footnote{Fixed point of the
running coupling $\lambda(t)$ occurs very close to the Planck
scale $t_{\rm P}$ with $\lambda(t_{\rm P})=0$.} In spite of all the objections, the remarkable conformity of the LHC tests and the Higgs mass range compatible with the CMB data makes this
model extremely attractive. In this section we shall discuss in some detail
certain aspects of the model of the Higgs inflation, following essentially the paper
\cite{Bar-Kam-Star2}.

The usual understanding of non-renormalizable theories is that
renormalization of higher-dimensional operators does not effect the
renormalizable sector of low-dimensional operators, because the former
ones are suppressed by powers of a cutoff -- the Planck mass $M_{\rm
  P}$ \cite{Weinberg}. Therefore, beta functions of the Standard Model
sector are not expected to be modified by gravitons.

The situation
with the non-minimal coupling is more subtle. Due to the mixing of the
Higgs scalar field with the longitudinal part of gravity in the
kinetic term of the Lagrangian (\ref{inf-grav}), an obvious
suppression of pure graviton loops by the effective Planck mass,
$M_{\rm P}^2+\xi\varphi^2\gg M_{\rm P}^2$, proliferates for large
$\xi$  to the sector of the Higgs field, so that certain parts of the
beta functions are strongly damped by a large $\xi$
\cite{our-ren,Wil}.
Therefore, a special combination of coupling constants
$\mbox{\boldmath$A$}$ which we call {\em anomalous scaling}
\cite{Bar-Kam1} becomes very small and lowers the CMB
compatible Higgs mass bound. The importance of this quantity
follows from the fact observed in \cite{Bar-Kam,Bar-Kam1,Bar-Kam2} that, due
to large $\xi$, quantum effects and their CMB manifestation are
universally determined by $\mbox{\boldmath$A$}$. The nature of
this quantity is as follows.

Let the model contain in addition to (\ref{inf-grav}) also a set of
scalar fields $\chi$, vector gauge bosons $A_\mu$ and spinors $\psi$,
which have an interaction with $\Phi$ dictated by the local gauge
invariance. If we denote by $\varphi$ the inflaton -- the only nonzero
component of the mean value of $\Phi$ in the cosmological state, then
the quantum effective action of the system takes a general form
\begin{eqnarray}
    S[g_{\mu\nu},\varphi]=\int d^{4}x\,g^{1/2}
    \left(-V(\varphi)+U(\varphi)\,R(g_{\mu\nu})
    -\tfrac12\,G(\varphi)\,(\nabla\varphi)^2+\ldots\right),
       \label{effaction}
    \end{eqnarray}
where $V(\varphi)$, $U(\varphi)$ and $G(\varphi)$ are the
coefficients of the derivative expansion, and we disregard the
contribution of higher-derivative operators that are negligible in the
slow-roll approximation of the inflation theory. In this
approximation, the dominant quantum contribution to these
coefficients comes from the heavy massive sector of the model. In
particular, the masses of the physical particles and Goldstone modes
$m(\varphi)$, generated by their quartic, gauge and Yukawa couplings
with $\varphi$, give rise to the Coleman--Weinberg potential -- the
one-loop contribution to the effective potential $V$ in
(\ref{effaction}). Since $m(\varphi)\sim\varphi$, for large
$\varphi$ this potential is given by the following sum of boson and
fermion contributions:
   \begin{eqnarray}
    &&V^{\rm 1-loop}(\varphi)=\sum_{
    \rm particles}
    (\pm 1)\,\frac{m^4(\varphi)}{64\pi^2}
    \,\ln\frac{m^2(\varphi)}{\mu^2}
    =\frac{\lambda\mbox{\boldmath$A$}}{128\pi^2}
    \,\varphi^4
    \ln\frac{\varphi^2}{\mu^2}+...  \label{Aviamasses}
    \end{eqnarray}
and thus determines the dimensionless coefficient
$\mbox{\boldmath$A$}$ -- the anomalous scaling associated with the
normalization scale $\mu$ in (\ref{Aviamasses}). Moreover, for $\xi\gg
1$ it is mainly this quantity and the dominant quantum correction to
$U(\varphi)$ \cite{Bar-Kam-Star1},
   \begin{eqnarray}
   U^{\rm 1-loop}(\varphi)=
    \frac{3\xi\lambda}{32\pi^2}\,\varphi^2
    \ln\frac{\varphi^2}{\mu^2}+...\, ,     \label{U1loop}
    \end{eqnarray}
which determine the quantum rolling force in the effective equation of the
inflationary dynamics \cite{Bar-Kam2,Bar-Nes} and  which yield the parameters of
the CMB generated during inflation \cite{Bar-Kam1}.

Inflation and its CMB are easy to analyse in the Einstein frame of
fields, denoted by $\hat g_{\mu\nu}$, $\hat\varphi$, in which the action $\hat
S[\hat g_{\mu\nu},\hat\varphi]=S[g_{\mu\nu},\varphi]$ has a minimal
coupling $\hat U=M_{\rm P}^2/2$, a canonically normalized inflaton
field $\hat G=1$, and the new inflaton potential $\hat{V}=M_{\rm P}^4
V(\varphi)/4U^2(\varphi)$.\footnote{The Einstein and Jordan frames
are related by the equations $\hat
g_{\mu\nu}=2U(\varphi)g_{\mu\nu}/M_{\rm P}^2$, $
(d\hat\varphi/d\varphi)^2=M_{\rm P}^2(GU+3U'^2)/2U^2$.} At the
inflationary scale with $\varphi>M_{\rm P}/\sqrt{\xi}\gg v$ and $\xi\gg
1$, this potential reads
        \begin{eqnarray}
        \hat{V}=\frac{\lambda
        M_{\rm P}^4}{4\,\xi^2}\,\left(1-\frac{2M_{\rm P}^2}{\xi\varphi^2}+
        \frac{\mbox{\boldmath$A_I$}}{16\pi^2}
        \ln\frac{\varphi}{\mu}\right),            \label{hatVbigphi}
        \end{eqnarray}
where the parameter $\mbox{\boldmath$A_I$}$ represents the anomalous
scaling  modified by the quantum correction to the
non-minimal curvature coupling (\ref{U1loop}),
       \begin{eqnarray}
        \mbox{\boldmath$A_I$}&=&\mbox{\boldmath$A$}-12\lambda=
        \frac3{8\lambda}\Big(2g^4 +
        \big(g^2 + g'^2\big)^2- 16y_t^4 \Big)-6\lambda.  \label{AI}
        \end{eqnarray}
This quantity -- which we shall call {\em inflationary anomalous
scaling} -- enters the expressions for the slow-roll parameters,
       \begin{eqnarray}
       \hat\varepsilon \equiv
       \frac{M_{\rm P}^2}2\left(\frac1{\hat V}\frac{d\hat
           V}{d\hat\varphi}\right)^2,\quad
       \hat\eta\equiv \frac{M_{\rm P}^2}{\hat V}
       \frac{d^2\hat V}{d\hat\varphi^2},
       \end{eqnarray}
and ultimately determines all the inflation characteristics. In
particular, the smallness of $\hat\varepsilon$ yields the range of the
inflationary stage $\varphi>\varphi_{\rm end}$, terminating at a value
of $\hat\varepsilon$ which we choose to be
$\hat\varepsilon_{\rm end}=3/4$.
Under the natural assumption that perturbation expansion is
applicable for $\mbox{\boldmath$A_I$}/64\pi^2\ll 1$,
the inflaton value at the exit
from inflation then equals $\varphi_{\rm end}\simeq 2M_{\rm P}/\sqrt{3\xi}$.
The value of
$\varphi$ at the beginning of the inflation stage of duration $N$ in
units of the e-folding number then reads \cite{Bar-Kam1}
    \begin{eqnarray}
    &&\varphi^2=\frac{4N}3\frac{M_{\rm
        P}^2}{\xi}\frac{e^x-1}x, \label{xversusvarphi}\\
    &&x\equiv\frac{N
    \mbox{\boldmath$A_I$}}{48\pi^2},           \label{x}
    \end{eqnarray}
where the special parameter $x$ directly involves the anomalous
scaling $\mbox{\boldmath$A_I$}$.

This relation determines the Fourier power spectrum for the scalar
metric perturbation $\zeta$,
$\Delta_{\zeta}^2(k) \equiv \langle k^3\zeta_{{\bf k}}^2\rangle
= \hat V/24\pi^2M_{\rm P}^4\hat\varepsilon$,
where the right-hand side is taken
at the  first horizon crossing, $k=aH$, relating the comoving
perturbation wavelength $k^{-1}$ to the e-folding number $N$,
    \begin{eqnarray}
    \Delta_{\zeta}^2(k)=
    \frac{N^2}{72\pi^2}\,\frac\lambda{\xi^2}\,
    \left(\frac{e^x-1}{x\,e^x}\right)^2.       \label{zeta}
    \end{eqnarray}
The CMB spectral index $n_s\equiv 1+d\ln\Delta_{\zeta}^2/d\ln
k=1-6\hat\varepsilon+2\hat\eta$ and the tensor to scalar ratio
$r=16\hat\varepsilon$ correspondingly read as\footnote{Note that for
$|x|\ll 1$ these predictions coincide with those of the $f(R)=(M_{\rm P}^2/2)(R+R^2/6M^2)$ inflationary model \cite{S80} with
the scalaron mass $M=M_{\rm P}\sqrt \lambda/\sqrt 3
\xi$ \cite{S83,Mukhanov_etal,S83a}.}
    \begin{eqnarray}
    &&n_s=
    1-\frac{2}{N}\, \frac{x}{e^x-1}~,           \label{ns}\\
    &&r=\frac{12}{N^2}\,
    \left(\frac{x e^x}{e^x-1}\right)^2~.          \label{r}
    \end{eqnarray}
Therefore, with the spectral index constraint $0.948 <n_s(k_0)<0.986$
(the combined WMAP+SPT+BAO+$H_0$ data at the $2\sigma$ confidence level
with the pivot point $k_0=0.002$
Mpc$^{-1}$ corresponding to $N\simeq 60$)  these
relations immediately give the range
$-12<\mbox{\boldmath$A_I$}<14$
for the inflationary anomalous scaling \cite{Bar-Kam1}.

In Standard Model $\mbox{\boldmath$A$}$ is
expressed in terms of the masses of the heaviest particles -- $W^\pm$
boson, $Z$ boson and top quark,
    \begin{eqnarray}
    m_W^2=\frac14\,g^2\,\varphi^2,\quad
    m_Z^2=\frac14\,(g^2+g'^2)\,\varphi^2,\quad
    m_t^2=\frac12\,y_t^2\,
    \varphi^2,                       \label{masses}
    \end{eqnarray}
and the mass of the three Goldstone modes
$m_G^2=V'(\varphi)/\varphi=\lambda(\varphi^2-v^2)\simeq
\lambda\varphi^2$. Here, $g$ and $g'$ are the $SU(2)\times U(1)$
gauge couplings, $g_s$ is the $SU(3)$ strong coupling, and $y_t$ is
the Yukawa coupling for the top quark. At the inflation stage, the
Goldstone mass-squared $m_G^2$ is non-vanishing, in contrast to its zero
on-shell value in the EW vacuum \cite{WeinbergQFT}. Therefore, Eq.
(\ref{Aviamasses}) gives the expression
   \begin{equation}
    {\mbox{\boldmath $A$}} =
    \frac3{8\lambda}\Big(2g^4 +
    \big(g^2 + g'^2\big)^2- 16y_t^4 \Big)+6\lambda.   \label{A0}
    \end{equation}
In the conventional range of the not yet observed Higgs mass 115 GeV$\leq M_{\rm H}\leq 180$ GeV (of the time of publication of \cite{particle}), this quantity was in  the range $-48<\mbox{\boldmath$A$}<-20$, which
contradicted the CMB range given above (though this contradiction was of $O(1)$ nature rather than of decimal orders of magnitude).

However, the RG running of coupling constants is strong enough and
drives ${\mbox{\boldmath $A$}}$  to the CMB compatible range at the
inflation scale. Below we show that the formalism of \cite{Bar-Kam1} stays
applicable but with the EW ${\mbox{\boldmath $A$}}$ replaced
by the running ${\mbox{\boldmath $A$}}(t)$, where $t=\ln(\varphi/\mu)$
is the running scale of the renormalization group (RG) improvement of
the effective potential \cite{ColemanWeinberg}.

\subsection{RG improvement}
According to the Coleman--Weinberg technique \cite{ColemanWeinberg},
the one-loop RG improved effective action has the form
(\ref{effaction}), with
    \begin{eqnarray}
    &&V(\varphi)=
    \frac{\lambda(t)}{4}\,Z^4(t)\,\varphi^4,  \label{RGeffpot}\\
    &&U(\varphi)=
    \frac12\Big(M_{\rm P}^2
    +\xi(t)\,Z^2(t)\,\varphi^{2}\Big),      \label{RGeffPlanck}\\
    &&G(\varphi)=Z^2(t).            \label{phirenorm1}
    \end{eqnarray}
Here, the running scale $t=\ln(\varphi/M_t)$ is normalized at the top quark mass $\mu=M_t$ (we denote physical (pole) masses by capital letters in contrast to running masses, see (\ref{masses}) above).\footnote{Application of the Coleman--Weinberg technique removes the ambiguity in the choice of the RG scale in cosmology -- an issue discussed in \cite{Woodard}.} The running couplings $\lambda(t)$,
$\xi(t)$ and the field renormalization $Z(t)$ incorporate a summation of
powers of logarithms and belong to the solution of the RG equations
    \begin{eqnarray}
    &&\frac{d g_i}{d t}
    =\beta_{g_i},\,\,\,\,\frac{dZ}{d t}
    =\gamma Z                   \label{renorm0}
    \end{eqnarray}
for the full set of coupling constants $g_i=(\lambda,\xi,g,g',g_s,y_t)$
in the ``heavy'' sector of the model with the corresponding beta functions $\beta_{g_i}$ and the anomalous dimension $\gamma$ of the Higgs field.

An important subtlety for these $\beta$ functions is the effect of the
non-minimal curvature coupling of the Higgs field. For large $\xi$, the
kinetic term of the tree-level action has a strong mixing between the
graviton $h_{\mu\nu}$ and the quantum part of the Higgs field $\sigma$
on the background $\varphi$. Symbolically, it has the structure
\[ (M_{\rm P}^2+\xi^2\varphi^2)h\nabla\nabla
h+\xi\varphi\sigma\nabla\nabla h+\sigma\triangle\sigma,\] which yields
a propagator whose elements are suppressed by a small $1/\xi$-factor
in all blocks of the $2\times2$ graviton-Higgs sector. For large
$\varphi\gg M_{\rm P}/\sqrt\xi$, the suppression of pure graviton
loops is, of course, obvious because  the effective Planck mass
squared strongly exceeds the Einstein one, $M_{\rm
  P}^2+\xi\varphi^2\gg M_{\rm P}^2$. Due to the mixing, this
suppression proliferates to the full graviton-Higgs sector of the
theory and yields the Higgs propagator
$s(\varphi)/(\triangle-m_H^2)$, which contains the suppression factor
$s(\varphi)$ given by
    \begin{eqnarray}
    s(\varphi)=
    \frac{M_{\rm P}^2+\xi\varphi^2}
    {M_{\rm P}^2+(6\xi+1)\xi\varphi^2}.         \label{s}
    \end{eqnarray}

This mechanism \cite{our-ren,Bar-Kam2,Bar-Nes} modifies the beta functions of
the SM sector \cite{Wil} at high energy scales because the factor
$s(\varphi)$, which is close to one at the EW scale $v\ll M_{\rm
  P}/\xi$, is very small for $\varphi\gg M_{\rm P}/\sqrt\xi$, $s\simeq
1/6\xi$. Such a modification justifies, in fact, the extension beyond
the scale $M_{\rm P}/\xi$ which is interpreted in \cite{BurgLeeTrott,Barbon} as
a natural validity cutoff of the theory.\footnote{\label{footnote4}The
smallness of this cutoff could be interpreted as an inefficiency of
the RG analysis beyond the range of validity of the
model. However, the cutoff $M_{\rm P}/\xi\ll M_{\rm P}$ of
\cite{BurgLeeTrott,Barbon} applies to energies (momenta) of scattering
processes in flat spacetime with a small EW value of $\varphi$. For
the inflation stage on the background of a large $\varphi$, this
cutoff gets modified due to the increase in the effective Planck mass
$M_{\rm P}^2+\xi\varphi^2\gg M_{\rm P}^2$
(and the associated decrease of the $s$-factor (\ref{s}) --
resummation of terms treated otherwise as perturbations in
\cite{BurgLeeTrott}). Thus the magnitude of the Higgs field at
inflation is not really indicative of the violation of the physical
cutoff bound.}

There is another important subtlety with the modification of beta
functions. Goldstone modes, in contrast to the
Higgs particle, do not have a mixing with gravitons in the kinetic term
\cite{Bez-Shap2}. Therefore, their contribution is not suppressed
by the $s$-factor of the above type. Separation of Goldstone
contributions from the Higgs contributions leads to the following
modification of the one-loop beta functions essentially
differing from those of \cite{Wil}:
    \begin{eqnarray}
    &&\beta_{\lambda} = \frac{\lambda}{16\pi^2}
    \left(18s^2\lambda
    +{\mbox{\boldmath $A$}}(t)\right)
    -4\gamma\lambda,                           \label{beta-lambda}\\
    &&\beta_{\xi} =
    \frac{6\xi}{16\pi^2}(1+s^2)\lambda
    -2\gamma\xi,                 \label{beta-xi}\\
    &&\beta_{y_t} = \frac{y_t}{16\pi^2}
    \left(-\frac{2}{3}g'^2
    - 8g_s^2 +\left(1+\frac{s}2\right)y_t^2\right)
    -\gamma y_t,                                    \label{beta-y}\\
    &&\beta_{g} = -\frac{39 - s}{12}
    \frac{g^3}{16\pi^2},                     \label{beta-g}\\
    &&\beta_{g'} =
    \frac{81 + s}{12} \frac{g'^3}{16\pi^2},  \label{beta-g1}\\
    &&\beta_{g_s} =
    -\frac{7 g_s^3}{16\pi^2}.                    \label{beta-gs}
    \end{eqnarray}
Here, the anomalous dimension $\gamma$ of the Higgs field  is given by
the standard expression in the Landau gauge,
    \begin{eqnarray}
    \gamma=\frac1{16\pi^2}\left(\,\frac{9g^2}4
    +\frac{3g'^2}4 -3y_t^2\right),                  \label{gamma}
    \end{eqnarray}
the anomalous scaling ${\mbox{\boldmath $A$}}(t)$ is defined by
(\ref{A0}),  and we have retained only the leading terms in $\xi\gg 1$. It
will be important in what follows that this anomalous scaling
contains the Goldstone contribution $6\lambda$, so that the full
$\beta_\lambda$ in (\ref{beta-lambda}) has a $\lambda^2$-term
unsuppressed by $s(\varphi)$ at large scale $t=\ln(\varphi/\mu)$.

The inflationary stage in units of Higgs field e-fold\-ings is very
short, which allows one to use an approximation linear in $\Delta
t\equiv t-t_{\rm end}= \ln(\varphi/\varphi_{\rm end})$, where the
initial data point is chosen at the end of inflation $t_{\rm end}$.
Therefore, for the beta functions (\ref{beta-lambda}) and
(\ref{beta-xi}) with $s\ll 1$ we have
    \begin{eqnarray}
    &&\lambda(t) = \lambda_{\rm end}\left(1
    - 4\gamma_{\rm end}\Delta t
    +\frac{\mbox{\boldmath $A$}(t_{\rm end})}{16\pi^2}\,
    \Delta t\right),                             \label{lambda-lin}\\
    &&\xi(t) = \xi_{\rm end}\Big(1
    -2\gamma_{\rm end}\Delta t
    +\frac{6\lambda_{\rm end}}{16\pi^2}\Delta t\Big),     \label{xi-lin}
    \end{eqnarray}
where $\lambda_{\rm end}$, $\gamma_{\rm end}$, $\xi_{\rm end}$ are
determined at $t_{\rm end}$, and ${\mbox{\boldmath $A$}}_{\rm
end}={\mbox{\boldmath $A$}}(t_{\rm end})$ is the particular value
of the running anomalous scaling (\ref{A0}) at the end of inflation.

On the other hand, the RG improvement of the effective action
(\ref{RGeffpot})--(\ref{phirenorm1}) implies that this action
coincides with the tree-level action  for a new field
$\tilde{\varphi}=Z(t)\varphi$ with running couplings as functions of
$t=\ln(\varphi/\mu)$ (the running of $Z(t)$ is slow and affects only
the multi-loop RG improvement). Then, in view of
(\ref{RGeffpot})--(\ref{RGeffPlanck}), the RG improved potential takes
at the inflation stage the form of the one-loop potential
(\ref{hatVbigphi}) for the field $\varphi$ with a particular choice of
the normalization point $\mu=\varphi_{\rm end}$ and all the couplings
replaced by their running values taken at $t_{\rm end}$. Therefore,
the formalism of \cite{Bar-Kam1} can be directly applied to find the CMB
parameters of the model, which now turn out to be determined by the
running anomalous scaling ${\mbox{\boldmath $A_I$}}(t)$ taken at
$t_{\rm end}$ .

In contrast to the inflationary stage, the post-infla\-tionary running
is very large and requires numerical simulation \cite{Bar-Kam-Star1}. We fix
the $t=0$ initial conditions for the RG equations (\ref{renorm0}) at
the top quark scale $M_t =171$ GeV. For the constants $g,g'$ and
$g_s$, they read \cite{particle}
    \begin{equation}
    g^2(0) = 0.4202,\  g'^2(0) = 0.1291,
    \ g_s^2(0) = 1.3460,                          \label{initial}
    \end{equation}
where $g^2(0)$ and $g'^2(0)$ are obtained by a simple one-loop RG flow
from the conventional values of $\alpha(M_Z)\equiv
g^2/4\pi=0.0338$, $\alpha'(M_Z)\equiv g'^2/4\pi=0.0102$ at
$M_Z$-scale, and the value $g_s^2(0)$ at $M_t$ is generated by the
numerical program of \cite{QCDfromZtotop}.
The analytical algorithm of transition between different scales for $g_s^2$
was presented in  \cite{DV,DV1,DV2}.
For the Higgs
self-interaction constant $\lambda$ and for the Yukawa top quark
interaction constant $y_t$, the initial conditions are determined by
the pole mass matching scheme originally developed in \cite{top,top1} and
presented in the Appendix of \cite{espinosa}.

The initial condition $\xi(0)$ follows from the CMB normalization
(\ref{zeta}), $\Delta_{\zeta}^2\simeq 2.5\times 10^{-9}$, at the
pivot point $k_0=0.002$ Mpc$^{-1}$, which we choose to
correspond to $N\simeq 60$. This yields the following estimate for
the ratio of coupling constants,
    \begin{equation}
\frac{1}{Z_{\rm in}^2}\frac{\lambda_{\rm in}}{\xi^2_{\rm in}}
    \simeq 0.5\times 10^{-9}
    \left(\frac{x_{\rm in}\,\exp x_{\rm in}}
    {\exp x_{\rm in}-1}\right)^2 ,             \label{final}
    \end{equation}
at the moment of the first horizon crossing for $N=60$, which we call
the ``beginning'' of inflation and label by $t_{\rm
in}=\ln(\varphi_{\rm in}/M_t)$ with $\varphi_{\rm in}$ defined by
(\ref{xversusvarphi}). Thus, the RG equations (\ref{renorm0}) for the
six couplings $(g,g',g_s,y_t,\lambda,\xi)$ with five initial
conditions and the final condition for $\xi$ uniquely determine the
needed RG flow.

The RG flow covers also the inflationary stage from the
chronological end of inflation $t_{\rm end}$  to $t_{\rm in}$. At the
end of inflation we choose the value of the slow roll
parameter $\hat\varepsilon=3/4$, and $\varphi_{\rm end}\equiv M_t
e^{t_{\rm end}}\simeq M_{\rm P}\sqrt{4/3\xi_{\rm end}}$. Thus, the
duration of inflation in units of inflaton field e-foldings
$t_{\rm in}-t_{\rm end}=\ln(\varphi_{\rm in}/\varphi_{\rm
end})\simeq\ln N/2\sim 2$ \cite{Bar-Kam-Star1} is very short relative to the
post-inflationary evolution $t_{\rm end}\sim 35$. The
approximation linear in the logarithms implies the bound $|{\mbox{\boldmath
$A_I$}}(t_{\rm end})|\Delta t/16\pi^2\ll 1$, which in view of
$\Delta t<t_{\rm in}-t_{\rm end}\simeq \ln N/2$ holds for
$|{\mbox{\boldmath $A_I$}}(t_{\rm end})|/16\pi^2$ $\ll 0.5$.

The running of ${\mbox{\boldmath $A$}}(t)$ depends strongly on the
behaviour of $\lambda(t)$. For small Higgs masses, the usual RG flow in
the SM leads to an instability of the EW vacuum caused by negative
values of $\lambda(t)$ in a certain range of $t$
\cite{Sher,espinosa}. The same happens in the presence of a
non-minimal curvature coupling.
\begin{figure}[h]
\includegraphics{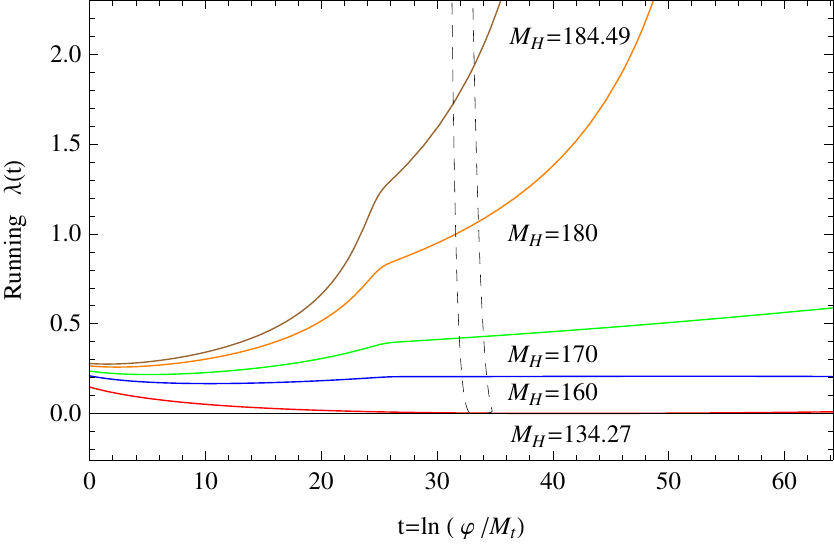}
\caption{Running $\lambda(t)$ for five values of the Higgs
mass above the instability threshold. Dashed curves mark the
boundaries of the inflation domain $t_{\rm end}\leq t\leq t_{\rm
in}$ \cite{Bar-Kam-Star1}.}
 \label{Fig.1}
\end{figure}
The numerical solution for $\lambda(t)$ is shown in Fig.\ref{Fig.1}
for five values of the Higgs mass and the value of top quark mass
$M_t=171$ GeV. The lowest one corresponds to the boundary of the
instability window,
    \begin{equation}
    M_{\rm H}^{\rm inst}\simeq 134.27\; {\rm GeV},      \label{criticalmass}
    \end{equation}
for which $\lambda(t)$ bounces back to positive values after
vanishing at $t_{\rm inst}\sim 41.6$ or $\varphi_{\rm inst}\sim 80
M_{\rm P}$. The shape of the corresponding effective potential in the
Einstein frame is depicted in Fig.~\ref{Fig.2} and shows the existence
of a false vacuum at this instability scale.
\begin{figure}[h]
\includegraphics{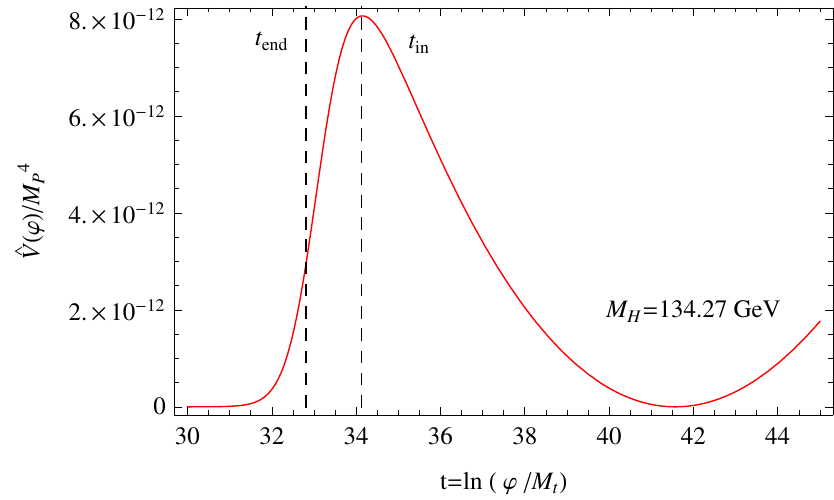}
\caption{\small The Einstein frame effective potential for the
instability threshold $M_{\rm H}^{\rm inst}=134.27$ GeV. A false
vacuum occurs at the instability scale $t_{\rm inst}\simeq 41.6$,
$\varphi_{\rm inst}\sim 80 M_{\rm P}$,
which is much higher than the Planck scale.
A possible domain of inflation (ruled out by the lower
$n_s$ CMB bound)
is again marked by dashed lines \cite{Bar-Kam-Star1}. }
\label{Fig.2}
\end{figure}
It turns out that the relevant $\xi(t)$ is nearly constant and is
about $5000$ (see below), so that the factor
(\ref{s}) at $t_{\rm inst}$ is very small, $s\simeq 1/6\xi\sim
0.00005$. Thus the situation is different from the usual SM
with $s=1$, and numerically the critical value turns out to be
higher than the known SM stability bound $\sim 125$ GeV
\cite{espinosa}.

Figure~1 shows that near the instability threshold $M_{\rm H}=M_{\rm
  H}^{\rm inst}$ the running coupling $\lambda(t)$ stays very small
for all scales $t$ relevant to the observable CMB. This follows from
the fact that the positive running of $\lambda(t)$
caused by the term $(18 s^2+6)\lambda^2$ in $\beta_\lambda$,
(see (\ref{beta-lambda})), is much slower for $s\ll 1$ than that of
the usual SM driven by the term $24\lambda^2$.

\begin{figure}[h]
\includegraphics{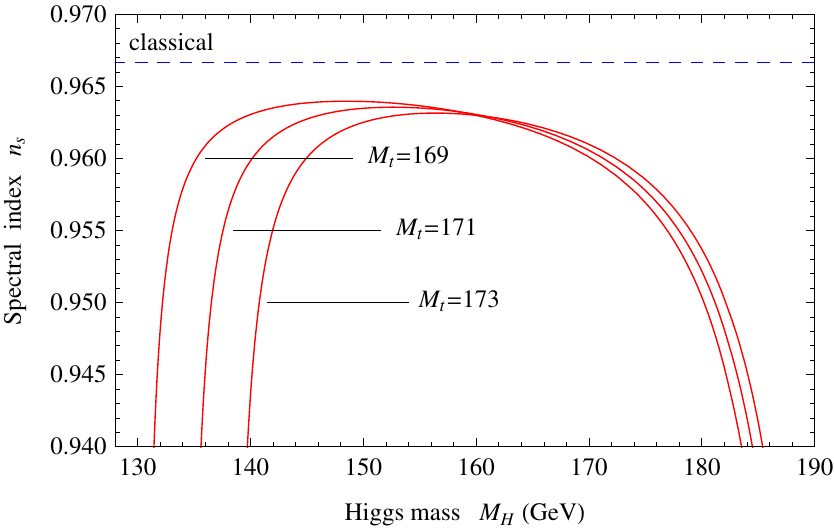}
\caption{ The spectral index $n_s$ as a function of the Higgs
mass $M_{\rm H}$ for three values  of the top quark mass \cite{Bar-Kam-Star1}.}
 \label{Fig.4}
\end{figure}

The RG running of ${\mbox{\boldmath $A_I$}}(t)$ explains the main
difference from the results of the one-loop calculations in
\cite{Bar-Kam1}. ${\mbox{\boldmath $A_I$}}(t)$ runs from big negative values
${\mbox{\boldmath $A_I$}}(0)<-20$ at the EW scale to small
but also negative values at the inflation
scale below $t_{\rm inst}$. This makes the CMB data compatible with
the generally accepted Higgs mass range. Indeed, the knowledge of the
RG flow immediately allows one to obtain
${\mbox{\boldmath$A_I$}}(t_{\rm end})$ and $x_{\rm end}$ and thus to
find the parameters of the CMB power spectrum (\ref{ns})--(\ref{r}) as
functions of $M_{\rm H}$. The parameter of primary interest --
the spectral index -- is given by (\ref{ns}) with $x=x_{\rm end}\equiv
N{\mbox{\boldmath
$A_I$}}(t_{\rm end})/48\pi^2$ and depicted in Fig.~\ref {Fig.4}. Even
for low values of the Higgs mass above the stability bound, $n_s$
falls into the range admissible by the CMB constraint.

The spectral index drops below 0.95 only for large $x_{\rm end}<0$
or large negative ${\mbox{\boldmath $A_I$}}(t_{\rm end})$, which
happens only when $M_{\rm H}$ either approaches the instability bound or
exceeds 180 GeV at the decreasing branch of the $n_s$ graph. Thus
we get lower and upper bounds on the Higgs mass, which both
follow from the lower bound of the CMB data. Numerical analysis for
the corresponding $x_{\rm end}\simeq -1.4$ gives for $M_t=171$ GeV
the following range for the  CMB compatible Higgs mass:
    \begin{equation}
    135.6\; {\rm GeV}\leq M_{\rm H}
    \leq 184.5\; {\rm GeV}.       \label{CMBmass}
    \end{equation}
Both bounds belong to the nonlinear domain of
(\ref{ns}) with $x_{\rm end}\simeq-1.4$, but the quantity
$|\mbox{\boldmath $A_I$}(t_{\rm end})|/16\pi^2=0.07\ll 0.5$
satisfies the restriction mentioned above, and their calculation is
still in the domain of our linear in logs approximation.

The upper bound on $n_s$ does not generate restrictions on $M_{\rm H}$.
The above
bounds were obtained for $M_t=171$ GeV. Results for the neighboring
values $M_t=171\pm2$ GeV are presented in Fig.~\ref{Fig.4} to show
the pattern of their dependence on $M_t$.

The expression (\ref{effaction}) is a truncation of the curvature
and derivative expansion of the full effective action. It was
repeatedly claimed that with large $\xi$ the weak field version of
this expansion on a flat (and empty) space background has a cutoff
$4\pi M_{\rm P}/\xi$ \cite{BurgLeeTrott,Barbon}. This scale is
essentially lower than the Higgs field during inflation $\varphi\sim
M_{\rm P}/\sqrt\xi$ and, therefore, seems to invalidate predictions based
on (\ref{effaction}) unless an unnatural suppression of
higher-dimensional operators is assumed. The attempt to improve the situation
by transition to the Einstein frame \cite{LernerMcDonald} was
claimed to fail \cite{BurgLeeTrott1,Hertzberg,Kaiser} for a
multiplet  Higgs field involving Nambu-Goldstone modes.

In the following, we show that these objections against naturalness are not
conclusive. First, as mentioned above, a large value of $\varphi$
during inflation is not really indicative of a large physical scale
of the problem. In contrast to curvature and energy density, the
inflaton itself is not a physical observable, but rather a
configuration space coordinate of the model. Second, we now show
that the inflation scale actually lies below the gradient expansion
cutoff, and this justifies the naturalness of the obtained results. No
transition to another conformal frame is needed for this purpose, but rather
a resummation accounting for a transition to a large $\varphi$
background.

Indeed, the main peculiarity of the model (\ref{inf-grav}) is that
in the background field method with small derivatives the role of
the effective Planck mass is played by $\sqrt{M_{\rm P}^2+\xi\varphi^2}$.
The power-counting method of \cite{BurgLeeTrott} underlying the
derivation of the cutoff $4\pi M_{\rm P}/\xi$ also applies here, but with
the Planck mass $M_{\rm P}$ replaced by the effective one,
$M_{\rm P}\to\sqrt{M_{\rm P}^2+\xi\varphi^2}>\sqrt\xi\varphi$. The resulting
cutoff is thus bounded from below by
    \begin{equation}
    \varLambda(\varphi)=\frac{4\pi\varphi}{\sqrt\xi},  \label{cutoff}
    \end{equation}
and this bound can be used as a {\em running} cutoff of the gradient
and curvature expansion. The origin of this cutoff can be
demonstrated in the one-loop approximation. When calculated in the
{\em Jordan} frame, for the one-loop divergences quadratic in the
curvature $R$ the dominating  $\xi$ contribution is (this can be
easily deduced from Appendix of \cite{Bar-Kam-Star1})
    \begin{equation}
    \xi^2\frac{R^2}{16\pi^2}.     \label{counterterm}
    \end{equation}
As compared to the tree-level part linear in the curvature $\sim
(M_{\rm P}^2+\xi\varphi^2)R$, the one loop $R^2$-term turns out to be
suppressed by the above cutoff factor
$16\pi^2(M_{\rm P}^2+\xi\varphi^2)/\xi^2\simeq\varLambda^2$.

The on-shell curvature estimate at the inflation stage reads
$R\sim V/U\sim\lambda\varphi^2/\xi$ in the Jordan frame,
so that the resulting curvature
expansion runs in powers of
    \begin{equation}
    \frac{R}{\varLambda^2}
    \sim\frac\lambda{16\pi^2}      \label{curvatureexpansion}
    \end{equation}
and remains valid in the usual perturbation theory range of SM, for which
$\lambda/16\pi^2\ll 1$. This works perfectly well in our Higgs
inflation model, because in the full CMB-compatible range of the
Higgs mass one has $\lambda<2$ (see Fig.~\ref{Fig.1}).

From the viewpoint of the gradient expansion for $\varphi$, this cutoff
is even more efficient. Indeed, the inflaton field gradient can be
expressed in terms of the inflaton potential $\hat V$ and the
inflation smallness parameter $\hat\varepsilon$ taken in the Einstein
frame,
$\dot\varphi\simeq(\varphi^2/M_{\rm P}^2)
(\xi\hat\varepsilon\hat{V}/18)^{1/2}$. With $\hat V\simeq\lambda
M_{\rm P}^4/4\xi^2$, this immediately yields the
gradient expansion in powers of
    \begin{equation}
    \frac{\partial}{\varLambda}
    \sim\frac1\varLambda\frac{\dot\varphi}\varphi\simeq
    \frac{\sqrt\lambda}{48\pi}
    \sqrt{2\hat\varepsilon},                \label{partialexpansion}
    \end{equation}
which is even better than (\ref{curvatureexpansion}) by a factor
ranging from $1/N$ at the beginning of inflation to $O(1)$ at the end
of it.

Equations (\ref{curvatureexpansion}) and (\ref{partialexpansion}) justify
the effective action truncation in (\ref{effaction}) in the
inflationary domain. Thus only multi-loop corrections to the
coefficient functions $V(\varphi)$, $U(\varphi)$, and $G(\varphi)$
may stay beyond control in the form of higher-dimensional
operators $(\varphi/\varLambda)^n$ and violate the flatness of the
effective potential necessary for inflation. However, in view of the
form of the running cutoff (\ref{cutoff}) they might be large, but
do not affect the shape of these coefficient functions because of
the field independence of the ratio $\varphi/\varLambda$. Only the
logarithmic running of couplings in
(\ref{RGeffpot})--(\ref{phirenorm1}) controlled by the RG dominates the
quantum input in the inflationary dynamics and its CMB
spectra.\footnote{This is like the logarithmic term in (\ref{hatVbigphi}),
which dominates over the nearly flat classical part of the inflaton
potential and qualitatively modifies the tree-level predictions of the
theory \cite{Bar-Kam-Star}.}

Before summing up, let us formulate once again the basic assumptions, made in \cite{Bar-Kam-Star,Bar-Kam-Star1,Bar-Kam-Star2}.
There was established the relation between the observable cosmological data (the spectral index $n_S$)
and the data coming from particle physics. This relation arises due to the fact that in the early
Universe the classical Friedmann evolution is essentially modified by quantum corrections. These
quantum corrections depend on interaction couplings of Standard Model particles with the Higgs field,
playing in the model under consideration the role of inflaton. To relate measured at the electroweak
scale values of these couplings with their hypothetical values at the inflationary scale we have used
the renormalization group formalism. We did it in spite of the well-known non-renormalizability
of quantum gravity, using two facts. First, below certain scale one can use an effective field theory
and all the participants of the related discussions agree that this scale is not lower than $m_P/\xi$.
Second, for large values of the scalar field (or, in other words, at high energies), the theory possesses
a scale invariance. It is this invariance which defends us from an uncontrollable growth of quantum
corrections. The question arises: is not  the transition between  these two ``safe'' regions of values
of the scalar field dangerous? The hypothesis which we used consists in the hope that the use of the continuous
$s$-factor constructs for us some kind of bridge between these two
regions, smoothly interpolating between low values of the Higgs field, where the effective theory is valid and high values,
where  the almost exact scale invariance is present.

\subsection{Palatini version of Higgs inflation model}
The nonminimal Higgs inflation model had a quite intensive and somewhat unexpected development in the years following the discovery of Higgs boson. Indeed, some old ideas put forward at the dawn of general relativity were applied in a quite new context. This is the so-called Palatini variational method \cite{Palatini,Ein-Pal,Francaviglia}. In the metric approach one has a relation between the affine connection and spacetime metric postulating
that the covariant derivative of the latter is vanishing. If torsion tensor is zero one obtains the Levi-Civita connection defined by Christoffel symbols. One can instead consider affine connection and metric as independent variables and try to obtain the relation between them starting from the variational principle. The Hilbert-Einstein action in this case can be obtained as follows: one defines the Riemann-Christoffel curvature tensor and Ricci tensor using only the affine connection, whereas for the construction of the Ricci scalar one needs to introduce metric. Then, varying  thus defined Hilbert-Einstein action with respect to the affine connection coefficients one obtains the equation giving the standard
Levi-Civita relation between the metric and connection, while the metric variation gives Einstein equations.

All this is true in the absence of a nontrivial coupling between matter fields and the quantities involving affine connection. In the presence of nonminimal coupling between Ricci scalar and the scalar field, the connection variation gives for the latter the expression containing beside the standard Chrisotoffel part extra terms depending on the scalar field. Indeed, if the Lagrangian of the model includes the nonminimal coupling term
 \begin{equation}
 L(g,\Gamma,\phi) = \sqrt{-g}U(\phi)g^{\mu\nu}R_{\mu\nu}(\Gamma),
 \label{Palatini}
 \end{equation}
then its variation with respect to the affine connection $\Gamma_{\mu\nu}^{\alpha}$ gives the expression containing $\phi$ and its spacetime derivatives,
\begin{equation}
\Gamma_{\mu\nu}^{\alpha} = \frac12g^{\alpha\beta}(g_{\beta\mu,\nu}+g_{\beta\nu,\mu}-g_{\mu\nu,\beta})
+\frac12\frac{U'(\phi)}{U(\phi)}(\delta_{\mu}^{\alpha}\phi_{,\nu}
+\delta_{\nu}^{\alpha}\phi_{,\mu}-g_{\mu\nu}g^{\alpha\beta}\phi_{,\beta}).
\label{Palatini1}
\end{equation}
Here the prime denotes the derivative with respect to the scalar field $\phi$. For a constant $U(\phi)$ additional terms in the right-hand side  of this equation vanish and it acquires standard metric form. For nonminimal coupling the components of Ricci tensor and scalar get nontrivial dependence on the scalar field and its spacetime derivatives, and the dynamics of the model changes significantly.

These effects were studied in detail for the nonminimal Higgs model in the paper \cite{Palatini-Higgs}. It was found that the two formalisms differ in their predictions for various cosmological parameters. The main reason is that the dependence on the nonminimal coupling parameter is very different in the two formalisms. For successful inflation, the Palatini approach prefers a much larger value of the nonminimal coupling than the metric approach. Unlike in metric formalism, in Palatini version inflaton stays well below the Planck scale therefor providing a natural inflationary epoch. In the subsequent paper \cite{Palatini-Higgs1} a special attention was paid to the question of unitarity.  It was stated that Higgs inflation does not suffer from unitarity violation since the UV cutoff lies parametrically much higher than the Hubble rate so that the unknown UV physics does not disrupt the inflationary dynamics. Higgs-Palatini inflation turns out to be UV-safe, minimal and endowed with predictive power.

The nonminimal Higgs inflation in the Palatini formalism with loop corrections was treated in \cite{Palatini-Higgs2}. It was shown that the  observable cosmological parameters are different in the Palatini approach and in the metric one. For example, the tensor-to-scalar ratio in the Palatini formalism is much lower than in the metric one. Thus, future observations can give indications concerning the nature of affine connection in our real physical Universe. In \cite{Palatini-Higgs3} the Coleman-Weinberg potential was studied for the nonminimal Higgs model within Palatini and metric formalisms and it was shown that these two formalisms predict different $e$-folding numbers. A wide class of scalar-tensor models of inflation in both formalisms was also considered in \cite{Palatini-Higgs4}. In particular, it was shown that for a simple Galileon model inflation naturally arises within Palatini formalism.

Quantum corrections in the Palatini version of the nonminimal Higgs model were studied and compared with the analogous effects in the metric theory in \cite{Palatini-Higgs5}. In particular, interesting relations between the bounds on top quark mass and the bounds on the parameter of the nonminimal coupling were discovered.

Finally, questions concerning different parametrizations of the Higgs doublets in the Palatini and metric formalisms were studied in the recent paper \cite{Palatini-Higgs6}. The Einstein-Cartan gravity with a nonvanishing torsion tensor was applied to Higgs inflation in \cite{Ein-Cart}. Numerous parameters of such a model make its comparison with observations much richer. On the other hand, the Einstein-Cartan theory is a simplest natural extension of the Palatini approach to gravity theory and, as stated in \cite{Ein-Cart}, there are no reasons to exclude it from the family of possible candidates on the role of a fundamental theory of gravity.

\section{Density matrix of the Universe and initial conditions for inflation}

As mentioned in Introduction, the success of Higgs inflation model does not extend to the solution of two big problems -- explanation of initial conditions for inflation and necessity to rely on more fundamental theory than the Standard Model of particles, which could be responsible for quantum nature of gravity at the energy scale characteristic of the quantum origin of inflationary Universe. Standard Model can be regarded as effective field theory representing low energy approximation to the theory of superstrings capable of probing the domain of Planckian scales, but suffering from the so-called landscape problem -- enormous ambiguity in the choice of its vacua \cite{land,land1}. At least naively, this problem might be regarded as a formulation of selection criteria that could restrict the energy domain of the physical realm of our Nature including such a phenomenon as quantum origin of the Universe. There is a hope that such a criterion can come from quantum cosmology within the formalism which would impose bounds on this energy scale, and if we are lucky enough this bound would be below the effective field theory cutoff so that we, in the absence of efficient nonperturbative methods, could proceed within perturbation theory. Here we present the attempt to implement this idea in the form of the density matrix of the Universe \cite{Bar-Kam-land, Bar-Kam-land1, Bar-Land, Bar-land1}.

\subsection{Microcanonical density matrix of the Universe}
The first step in pursuit of the above program consists in the denial of a pure nature of the quantum state of the Universe and in its replacement by a density matrix. Among the reasons for this denial is that the choice of a certain quantum state, like the no-boundary or tunneling one, is always associated with additional assumptions. So, in the spirit of Occam razor denying redundant assumptions, there is a room for being democratic and try  handling all possible quantum states on equal footing, which in fact implies transition to their mixture -- a density matrix. The very idea that instead of a pure quantum state one can consider a density matrix in the form of the Euclidean quantum gravity path integral was pioneered by D. Page in \cite{Page}. However, Euclidean quantum gravity which, in particular, underlies the Hartle-Hawking no-boundary wavefunction can hardly be a fundamental concept. Rather, it should be derived as a calculational tool from canonical quantization in real physical spacetime \cite{Bar}, just like it works in conventional quantum field theory. Such a derivation might look as follows.

Consider instead of one concrete solution $|\,\Psi\,\rangle$ of the Wheeler-DeWitt equation (\ref{WDW}) a full set of them and construct the density matrix according to the following transition
    \begin{eqnarray}
    |\,\Psi\,\rangle \to \hat\rho=\sum\limits_{{\rm all}\,|\,\Psi\,\rangle}|\,\Psi\,\rangle
    \langle\,\Psi\,|,                        \label{psi_to_rho}
    \end{eqnarray}
where summation runs over all such solutions, $\hat{\cal H}|\,\Psi\,\rangle=0$. Unfortunately, we do not yet know a precise operator realization of the set of the quantum Hamiltonian and momentum constraints in gravity theory $\hat{\cal H}=\{\hat H_\perp({\bf x}),\hat H_a({\bf x})\}\equiv\hat H_\mu$, which we will collectively denote by $\hat H_\mu$ with the condensed index $\mu=(\perp\!{\bf x},a{\bf x})$ including the continuous spatial coordinate label. Therefore it is hard to define the set of conditions functionally restricting this space of solutions, but one can formally proceed with the definition (\ref{psi_to_rho}) treating it as a projector in terms of generalized operator-valued delta-function,
    \begin{eqnarray}
    &&\hat\rho=\frac1Z\,\delta(\hat{\cal H}),
    \quad Z={\rm tr}\,\delta(\hat{\cal H}),    \label{rho}\\
    &&\delta(\hat{\cal H})=\vphantom{\bigg[\prod_\mu\bigg]}^{{}^{\bf\textstyle\prime\prime}}
    \prod_\mu \delta(\hat H_\mu)\vphantom{\bigg[\prod_\mu\bigg]}^{{}^{\bf\textstyle\prime\prime}}.   \label{projector}
    \end{eqnarray}
Quotation marks indicate here that the product of non-commuting operators $\hat H_\mu$ should not be understood literally, but rather interpreted as an operatorial generalized function projecting onto the kernel of all the constraints, that is satisfying the equation $\hat H_\mu \delta(\hat{\cal H})=\delta(\hat{\cal H}) \hat H_\mu=0$. Consistency of the definition of this projector follows from the involution algebra of these constraints -- their commutator being their linear combination \cite{Bar-land1}. The commutator algebra of constraints does not form the Lie algebra of a group, so that it is impossible to define such a projector by integration over the group volume.

However, the projector (\ref{projector}) can be constructed along the lines of the BFV quantization \cite{BFV,BFV1} in the form of the path integral over BRST-extended set of fields \cite{Bar-land1}. Its matrix element in the functional coordinate representation of 3-metric coefficients and matter fields $q=(g_{ab}({\bf x}),\phi({\bf x}))$ turns out to be given by the canonical Faddeev-Popov path integral,
    \begin{eqnarray}
    \langle\,q_+|\,\delta(\hat{\cal H})\,|\,q_-\rangle=
    \!\!\!\!\!
    \int\limits_{\,\,\,\,\,\,
    q(t_\pm)=\,q_\pm}
    \!\!\!\!\!\!\!\!\!
    D[\,q,p,N\,]\;
    \exp\left[\,i\!
    \int_{\,\,t_-}^{t_+} dt\,
    (p\,\dot q-N^\mu H_\mu)\,\right],   \label{rhocanonical}
    \end{eqnarray}
with the canonical ADM form of the gravitational action \cite{ADM} and the gauge fixing procedure encoded in the integration measure $D[\,q,p,N\,]$. This is the integration over histories of gravitational and matter phase space variables $(q,p)$ and Lagrange multipliers $N^\mu$ which are nothing but the ADM lapse and shift functions $N^\mu=(N^\perp({\bf x}), N^a({\bf x}))$. These histories interpolate between the arguments $q_\pm$ of the projector kernel \cite{Bar}.

Note that the parameter $t$ here, which looks exactly like a physical time variable in the canonical formalism of general relativity, originally arose as an operator ordering parameter that allows one to account for non-Abelian nature of quantum constraints. Originally its role was to extend to the operator level with non-commuting $\hat H_\mu$ the c-number delta function $\int dN\,\exp(-iN^\mu H_\mu)=\prod_\mu\delta(H_\mu)$ in the representation of the Fourier integral over $N^\mu$ (single-time not the path-integral one).

Physical interpretation of the definition (\ref{rho}) is that it is the analogue of microcanonical density matrix for the usual non-gauge dynamical system with a fixed value of the total conserved energy $E$, $\hat\rho=\delta(\hat H-E)/Z$ -- equipartition of all physical states with one and the same energy value. In spatially closed cosmology there are no global conserved charges including its energy which is not being defined at all. Instead, the closed cosmology has local conserved objects -- the constraints $H_\mu$ -- whose only physically meaningful and mathematically consistent value is zero. Therefore, the density matrix (\ref{rho}) can be regarded as a definition of microcanonical ensemble in quantum cosmology -- a universal equipartition of all physical states satisfying the condition of local diffeomorphism invariance.

The partition function of this ensemble,
    \begin{eqnarray}
    Z&=&\int dq\,\mu(q)\,\langle\,q_+|\,\delta(\hat{\cal H})\,|\,q_-\rangle\,\Big|_{\,q_\pm=q}\nonumber\\
    &=&
    \int\limits_{\,\,\,
    \rm periodic}
    \!\!\!\!\!
    D[\,q,p,N\,]\;
    \exp\left[\,i\!
    \int_{\,\,t_-}^{t_+} dt\,
    (p\,\dot q-N^\mu H_\mu)\,\right],   \label{partition}
    \end{eqnarray}
obviously implies path integration over closed periodic histories (the measure of the physical inner product $\mu(q)$ being absorbed in the path integration measure $D[\,q,p,N\,]$ of such histories \cite{Bar-land1}). Note that despite a usually exploited relation between microcanonical thermodynamics and imaginary time formalism the path integral here is still over fields in Lorentzian signature spacetime.

The further transformation of the path integral consists in Gaussian integration over momenta and the use of relativistic gauge conditions. This eventually allows one to assemble the initially (3+1)-splitted fields $q=\big(g_{ab}(t,{\bf x}),\phi(t,{\bf x})\big)$ and $N^\mu=\big(N(t,{\bf x}),N^a(t,{\bf x})\big)$ into covariant multiplets of spacetime metric $g_{\mu\nu}(x)$ and matter $\phi(x)$ variables and rewrite the partition function as a covariant Lagrangian path integral over periodic fields on the manifold with a variable $x^0=t$ compactified to a circle $S^1$,
    \begin{eqnarray}
    &&Z=
    \!\!\int\limits_{\,\,\rm periodic}
    \!\!\!\! D[\,g_{\mu\nu},\phi\,]\;
    e^{iS[\,g_{\mu\nu},\phi\,]}.         \label{Z}
    \end{eqnarray}

The notion of imaginary time or Euclidean spacetime arises only at the level of semiclassical expansion. This is a usually accepted assumption that the integrand of a path integral has two types of analyticity -- one with respect to the extension of integration fields to their complex values and another one is with respect to transition of their spacetime arguments into the complex plane of time.\footnote{In gravity theory the second type of analyticity can be associated with the first type when the Euclidean spacetime is attained not at imaginary values of time parameter, but at the imaginary value of the lapse function.} Therefore, if the stationary point of the path integral (\ref{Z}) is not available, that is no periodic solutions of equations of motion for $S[\,g_{\mu\nu},\phi\,]$ exist in Lorentzian signature spacetime, then such periodic solutions can be looked for in imaginary time. And if they exist, they can serve as saddle point configurations in Euclidean spacetime with the negative of the Euclidean action $-S_{\rm E}=iS$ in the exponential of (\ref{Z}).

\begin{figure}
\vspace{-0.7cm}
\centerline{\includegraphics[height=.21\textheight]{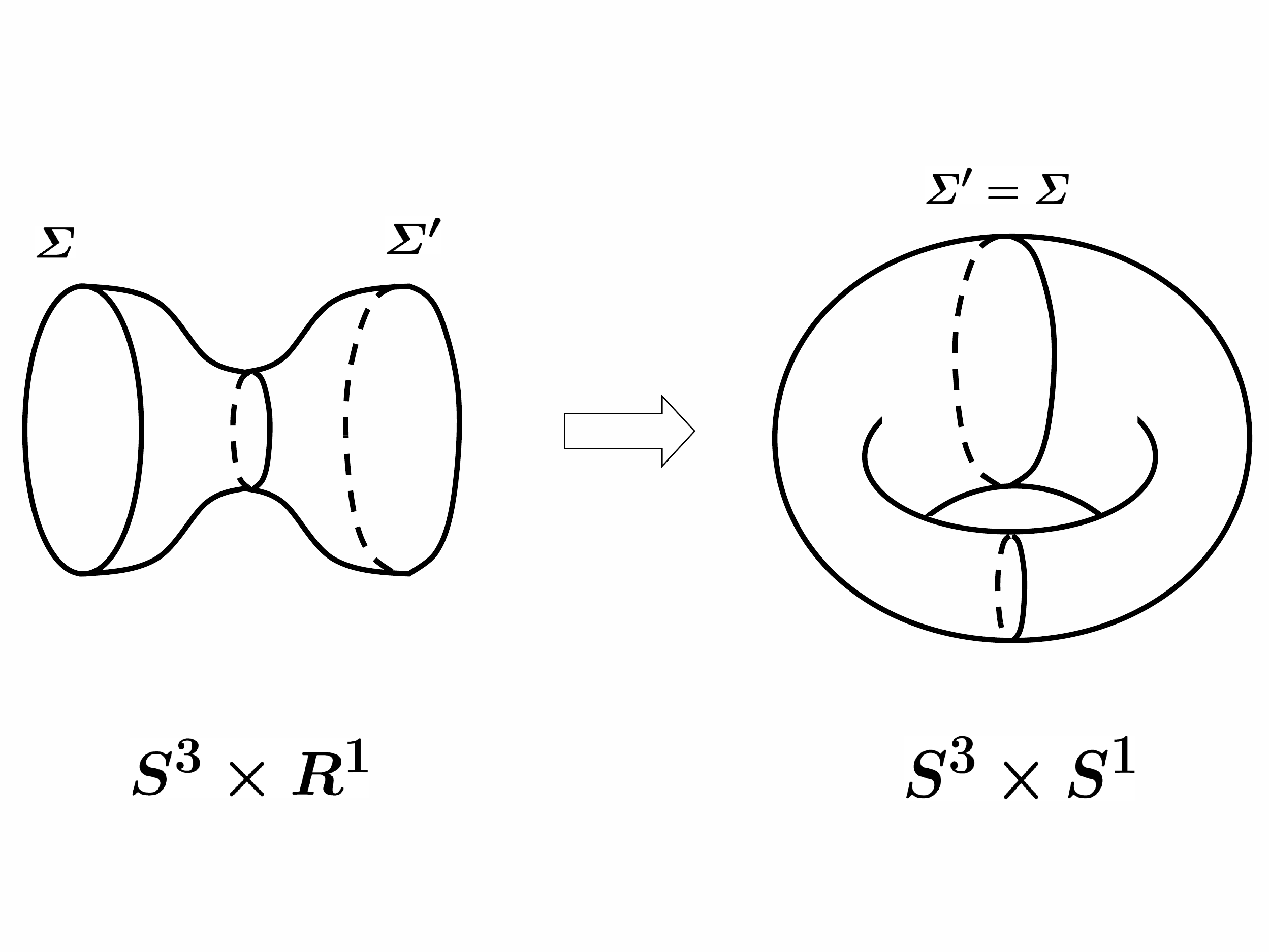}}
\vspace{-0.4cm}
\caption{\small Origin of the partition function instanton from the density matrix instanton by the procedure of gluing the boundaries $\Sigma$ and $\Sigma'$ -- tracing the density matrix.}
 \label{Fig.10}
\end{figure}

\begin{figure}
\centerline{\includegraphics[height=.12\textheight]{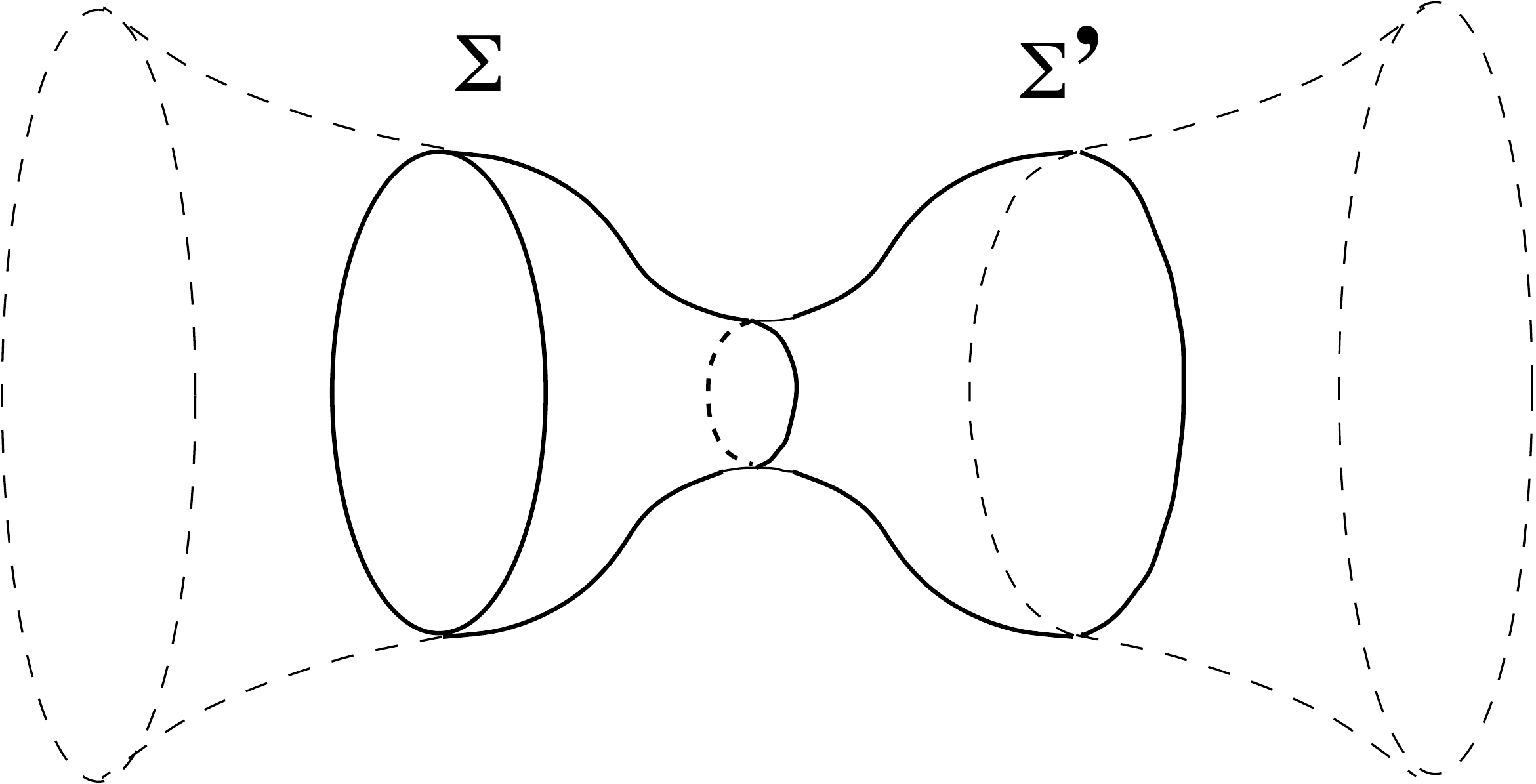}}
\caption{Picture of instanton representing the density matrix. Dashed lines
depict the Lorentzian Universe nucleating from the instanton at the
minimal surfaces $\Sigma$ and $\Sigma^{^{\textstyle\bf,}}$.}
\label{Fig.11}
\end{figure}

Such a solution in Euclidean spacetime, or cosmological instanton, for spatially closed cosmology with $S^3$ topology of spatial sections will have the 4-dimensional topology $S^3\times S^1$. Its origin from the corresponding instanton for the density matrix with the topology $S^3\times R^1$ is graphically represented on Fig.\ref{Fig.10} as a result of glueing two spatial hypersurfaces $\Sigma$ and $\Sigma'$ associated with two arguments of the density matrix (\ref{rho}).
At the glued minimal surfaces $\Sigma$ and $\Sigma'$, labelled respectively by the values $\tau$ and $\tau'$ of Euclidean time, all fields have vanishing time derivatives, which follows from matching periodicity conditions. Therefore, a real valued instanton solution can be analytically continued to the imaginary axes of the Euclidean time (or real axes of the Lorentzian time $t$), $\tau\to\tau+it$ and $\tau'\to\tau'+it'$, and the resulting solutions as functions of $t$ and $t'$ will also be real valued. The combined Euclidean-Lorentzian instanton is then graphically represented on Fig.\ref{Fig.11} where its Lorentzian part is depicted by dashed lines.

\begin{figure}
\centerline{\includegraphics[height=.13\textheight]{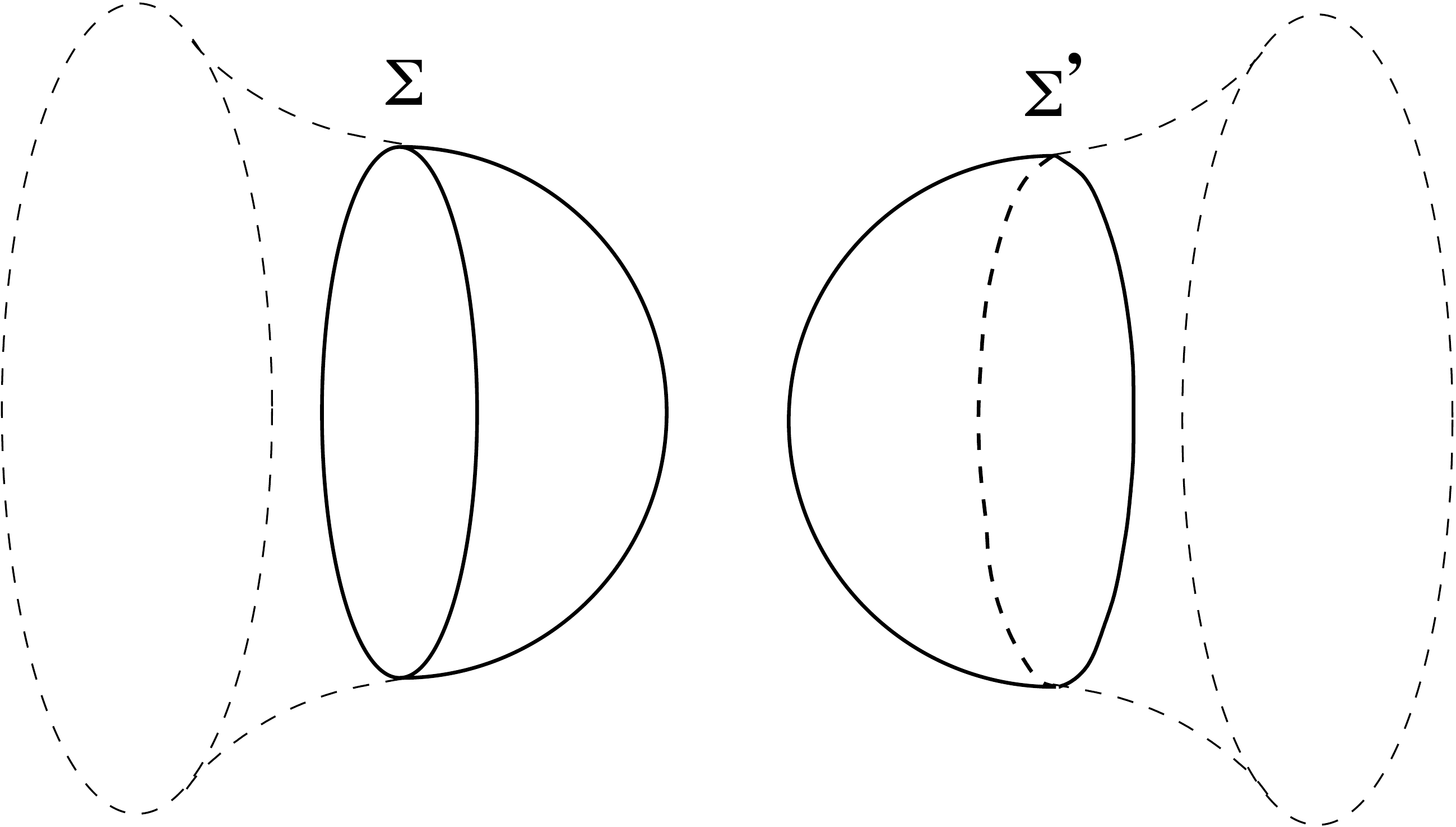}}
\caption{Density matrix of the pure Hartle-Hawking state represented by the
union of two no-boundary vacuum instantons.}
\label{Fig.12}
\end{figure}

The last picture demonstrates the difference between the pure no-boundary state from the density matrix one. The no-boundary state, which originally is defined by the Euclidean quantum gravity path integral, can also be represented by the density matrix having a factorisable form and corresponding to the instanton shown on Fig.\ref{Fig.12}. What distinguishes it from the impure state described by the density matrix is that its underlying instanton is the union of two disjoint manifolds. The instanton bridge between the surfaces $\Sigma$ and $\Sigma^{^{\textstyle\bf,}}$ of Fig.\ref{Fig.11} is broken, which implies absence of correlations between the quantum states of fields on these surfaces.

Thus, the starting point of the above construction  \cite{Bar-Kam-land,Bar-Kam-land1,Bar-Land}  is the density matrix $\hat\rho$ with two surfaces carrying its field arguments. These surfaces {\em semiclassically} are the boundaries of either Euclidean or Lorentzian spacetime, depending on the relevant size of the scale factor. The entire saddle point solution for $\hat\rho$ consists respectively of the Euclidean spacetime interpolating between them or of the Euclidean spacetime between $\Sigma$ and $\Sigma'$, sandwiched between the two layers of the Lorentzian spacetime. These two layers interpolate from $\Sigma$ to the unprimed argument of the density matrix and from $\Sigma'$ to its primed argument and correspond in the density matrix to the chronological and anti-chronological evolution factors of the well-known Schwinger-Keldysh technique for expectation values \cite{Schwinger-Keldysh1,Schwinger-Keldysh2}. When calculating the statistical sum trace these two factors cancel out in view of unitarity, and the only contribution that remains comes from the Euclidean domain between the Euclidean-Lorentzian transition surfaces $\Sigma$ and $\Sigma'$. These surfaces are uniquely determined from the condition of smooth periodicity in the Euclidean time on the compact $S^1$, or as two turning points of the Euclidean trajectory for $a(\tau)$.

\subsection{Effective action and cosmological bootstrap in CFT driven cosmology}
Productive application of the above construction can be demonstrated in the model of inflation driven by conformal field theory. This is the Einstein gravity model with a primordial cosmological constant $\varLambda$ and the matter sector dominated by a large number of fields $\Phi$ conformally coupled to gravity -- conformal field theory (CFT) with the action $S_{CFT}[\,g_{\mu\nu},\varPhi\,]$,
    \begin{eqnarray}
    &&S_{\rm E}[\,g_{\mu\nu},\varPhi\,]=-\frac{M_P^2}2
    \int d^4x\,g^{1/2}\,(R-2\varLambda)
    +S_{CFT}[\,g_{\mu\nu},\varPhi\,].    \label{tree0}
    \end{eqnarray}
Important property which allows one to overstep the limits of the usual semiclassical expansion consists here in the possibility to omit integration over conformally non-invariant matter fields and spatially-inhomogeneous metric modes on top of a dominant contribution of these conformal species. Integrating them out one obtains the effective gravitational action $S_{\rm eff}[\,g_{\mu\nu}]$ which differs from (\ref{tree0}) by $S_{CFT}[\,g_{\mu\nu},\varPhi\,]$ replaced with $\varGamma_{CFT}[\,g_{\mu\nu}]$ -- the effective action of $\varPhi$ on the background of $g_{\mu\nu}$,
    \begin{eqnarray}
    e^{-\varGamma_{C\!F\!T}[\,g_{\mu\nu}]}=\int D\varPhi\,
    e^{-S_{C\!F\!T}[\,g_{\mu\nu},\varPhi\,]}. \label{GammaCFT}
    \end{eqnarray}
On Friedmann-Robertson-Walker (FRW) background this action is exactly calculable by using the local conformal transformation to the static Einstein universe and well-known local trace anomaly. The resulting  $\varGamma_{CFT}[\,g_{\mu\nu}]$ turns out to be the sum of the anomaly contribution and free energy of conformal matter fields at the effective temperature determined by the circumference of the compactified time dimension $S^1$. This is an important calculational advantage provided by local Weyl invariance of quantum matter.

Physics of the CFT driven cosmology is entirely determined by this effective action. Solutions of its equations of motion, which give a dominant contribution to the statistical sum, are the cosmological instantons of $S^1\times S^3$ topology, which have the Friedmann-Robertson-Walker metric
    \begin{eqnarray}
    ds^2=N^2(\tau)\,d\tau^2
    +a^2(\tau)\,d^2\Omega^{(3)}              \label{FRW}
    \end{eqnarray}
with a periodic lapse function $N(\tau)$ and scale factor $a(\tau)$ -- functions of the Euclidean time belonging to the circle $S^1$ \cite{Bar-Kam-land}. These instantons can be interpreted as initial conditions for the cosmological evolution $a_L(t)$ in the physical Lorentzian spacetime. This evolution follows from $a(\tau)$ by analytic continuation to real time $t$, $a_L(t)=a(\tau_*+it)$, at the point of the maximum value of the Euclidean scale factor $a_+=a(\tau_*)$. Looking ahead to the formulated above goals of this setup, let us say that these instantons will exist only in the finite range of $\varLambda$. Under the assumption that $\varLambda$ is a function of the inflaton field staying, say, in slow roll regime and exiting it according to conventional inflation scenario, this restriction would mean the formation of initial conditions of inflation. On the other hand, this restriction may be interpreted as the selection criterion in the landscape of stringy vacua.

The realization of this program is as follows. For cosmology with the metric (\ref{FRW}) its effective action reads \cite{Bar-Kam-land}
    \begin{eqnarray}
    &&S_{\rm eff}[\,a,N\,]=6\pi^2 M_P^2\int_{S^1} d\tau\,N \left\{-aa'^2
    -a+ \frac\varLambda3 a^3\right.\nonumber \\
    &&\left.\qquad\qquad\qquad\qquad\quad+B\,\Big(\frac{a'^2}{a}
    -\frac{a'^4}{6 a}\Big)
    +\frac{B}{2a}\,\right\}+F(\eta),               \label{effaction0}\\
    &&F(\eta)=\sum_{\omega}(\pm1)\ln\big(1\mp
    e^{-\omega\eta}\big),\quad \eta=\int_{S^1} \frac{d\tau N}a,     \label{period}
    \end{eqnarray}
where $a'\equiv da/Nd\tau$. The first three terms in curly brackets of (\ref{effaction0}) represent the Einstein action with a fundamental cosmological constant $\varLambda\equiv 3H^2$ ($H$ is the corresponding Hubble parameter). The constant $B$ is a coefficient of the contributions of the conformal anomaly and vacuum (Casimir) energy $(B/2a)$ on a conformally related static Einstein spacetime mentioned in Introduction. This constant,
    \begin{eqnarray}
    B=\frac{\beta}{8\pi^2 M_P^2},         \label{B}
    \end{eqnarray}
expresses via the coefficient $\beta$ of the Gauss-Bonnet term $E=R_{\mu\nu\alpha\gamma}^2-4R_{\mu\nu}^2+ R^2$ in the trace anomaly of conformal matter fields
    \begin{eqnarray}
    g_{\mu\nu}\frac{\delta
    \Gamma_{\rm CFT}}{\delta g_{\mu\nu}} =
    \frac{1}{4(4\pi)^2}g^{1/2}
    \left(\alpha \nabla^2R +
    \beta E +
    \gamma C_{\mu\nu\alpha\beta}^2\right). \label{anomaly}
    \end{eqnarray}

The effective action is independent of the anomaly coefficients $\alpha$ and $\gamma$, because Weyl tensor $C_{\mu\nu\alpha\beta}$ identically vanishes for any Friedmann metric and it is assumed that $\alpha$ can be renormalized to zero by a local counterterm $\sim\int d^4x\,g^{1/2}R^2$. This, in particular, guarantees absence of higher derivative terms in (\ref{effaction0}) -- non-ghost nature of the scale factor -- and simultaneously endows the renormalized Casimir energy with a special value proportional to $B/2=\beta/16\pi^2M_P^2$ \cite{Bar-Kam-land}. Both of these properties are critically important for the instanton solutions of effective equations.

Finally, $F(\eta)$ in (\ref{period}) is the free energy of conformal fields on static Einstein universe space (to which the conformal rescaling mentioned above was done) -- a typical boson or fermion sum over CFT field oscillators with energies $\omega$ on a unit 3-sphere, $\eta$ playing the role of the inverse temperature --- an overall circumference of $S^1$ in the $S^1\times S^3$ instanton geometry, which is calculated in units of the conformal time (\ref{period}).

The statistical sum (\ref{Z}) is dominated by the solutions of the effective equation, $\delta S_{\rm eff}/\delta N(\tau)=0$, which in the cosmic time gauge ($N=1$) reads
    \begin{eqnarray}
    &&-\frac{\dot a^2}{a^2}+\frac{1}{a^2}
    -B \left(\,\frac{\dot a^4}{2a^4}
    -\frac{\dot a^2}{a^4}\right) =
    \frac\varLambda3+\frac{C}{ a^4},\ \dot a=\frac{da}{d\tau},               \label{efeq0}\\
    &&C =
    \frac{B}2+\frac1{6\pi^2 M_P^2}\,\sum_\omega\frac{\omega}
    {e^{\omega\eta}\mp 1}. \label{bootstrap}
    \end{eqnarray}
This is the modification of the Euclidean Friedmann equation by the conformal anomaly term $\sim B$, the radiation energy term $C/a^4$ which is the sum of the Casimir energy $\sim B/2$ and the energy of thermally excited particles with the inverse temperature $\eta$ given by (\ref{period}). Note that the constant $C$ is a nonlocal functional of the history $a(\tau)$ -- Eq. (\ref{bootstrap}) plays the role of the {\em bootstrap} equation for the amount of radiation determined by the background on top of which this radiation evolves and produces back reaction.

Eq. (\ref{efeq0}) can be solved for $\dot a^2$,
    \begin{eqnarray}
    \dot{a}^2 &=& \sqrt{\frac{(a^2-B)^2}{B^2}
    +\frac{2H^2}{B}\,(a_+^2-a^2)(a^2-a_-^2)}-\frac{a^2-B}{B},            \label{mainEq}\\
    a_\pm^2&\equiv&
    \frac{1\pm\sqrt{1-4CH^2}}{2H^2},          \label{apm}
    \end{eqnarray}
to give a periodic oscillation of $a$ between its maximal and minimal values $a_\pm$, provided that at $a_-$ we have a turning point with a vanishing $\dot a$, which means that $a_-^2>B$. This inequality together with the requirement of reality of turning points $a_\pm$ immediately yields the following restrictions on the range of $H^2$ and $C$
    \begin{eqnarray}
    &&H^2\leq\frac1{2B},
    \quad \frac1{4H^2}\geq C\geq B-B^2H^2.     \label{bounds}
    \end{eqnarray}

\subsection{Garland instantons and elimination of infrared catastrophe}
The solutions of this integro-differential equation give rise to the set of periodic $S^3\times S^1$ instantons with the oscillating scale factor -- {\em garlands} -- that can be regarded as the thermal alternative to the Hartle-Hawking instantons \cite{Bar-Kam-land,Bar-Land,Bar-Kam-Def1}. The scale factor oscillates $m$ times ($m=1,2,3,...$) between the maximum and minimum values (\ref{apm}), $a_-\leq a(\tau)\leq a_+$, so that the full period of the conformal time (\ref{period}) is the $2m$-multiple of the integral between the two neighboring turning points of $a(\tau)$, $\dot a(\tau_\pm)=0$,
    \begin{eqnarray}
    &&\eta=2m\int_{a_-}^{a_+}
    \frac{da}{\dot a\,a}.                       \label{period1}
    \end{eqnarray}
This value of $\eta$ is finite and determines effective temperature $T=1/\eta$ as a function of $G=1/8\pi M_P^2$ and $\varLambda=3H^2$. This is the artifact of a microcanonical ensemble in cosmology with only two freely specifiable dimensional parameters --- the gravitational and cosmological constants.

\begin{figure}
\centerline{\includegraphics[height=.4\textheight]{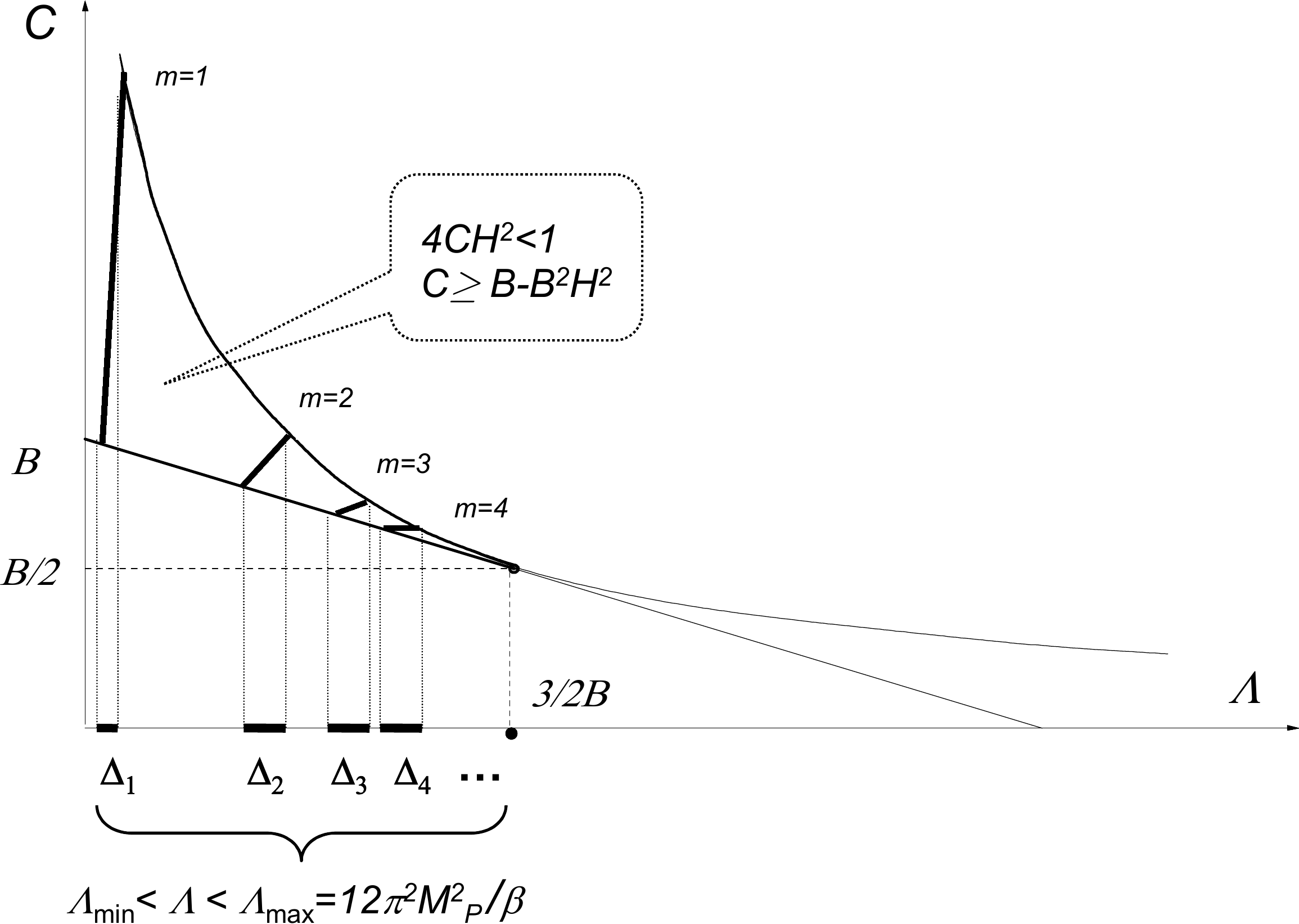}}
\caption{\small Band range of garland instantons formed by projections $\Delta_m$ of their $m$-folded one-parameter families onto the axis of $\varLambda=3H^2$.}
 \label{Fig}
\end{figure}

According to the bounds (\ref{bounds}) these garland-type instantons exist only in the limited range of the cosmological constant $\varLambda=3H^2$ \cite{Bar-Kam-land}. They belong to the domain in the two-dimensional plane of the Hubble constant $H^2$ and the amount of radiation constant $C$. In this domain they form an countable, $m=0,1,2,...$, sequence of one-parameter families -- curves interpolating between the lower straight line boundary $C=B-B^2H^2$ and the upper hyperbolic boundary $C=1/4H^2$. Each curve corresponds to respective $m$-folded instantons of the above type. Therefore, the range of admissible values of $\varLambda$,
    \begin{eqnarray}
    \varLambda_{\rm min}\leq\varLambda\leq
    \varLambda_{\rm max}=
    \frac{12\pi^2M_P^2}\beta=\frac3{2B},    \label{Lambda_range}
    \end{eqnarray}
has a band structure, each band $\Delta_m$ being a projection of the $m$-th curve to the $\varLambda$ axis. The sequence of bands of ever narrowing widths with $m\to\infty$ accumulates at the upper bound of this range $H^2_{\rm max}=1/2B$. The lower bound $H^2_{\rm min}$ -- the lowest point of $m=1$ family -- can be obtained numerically for any field content of the model.

Another set of solutions follows from rewriting the effective equation (\ref{efeq0}) in the form (retaining in contrast to (\ref{mainEq}) both signs of the square root, which is possible in the other range of $a^2$, $a^2<B$)
    \begin{eqnarray}
    \dot{a}^2 =1-\frac{a^2}{B}\left(\,1
    \pm\sqrt{1-2BH^2-\frac{B(2C-B)}{a^4}}\,\right) \label{efeq2}
    \end{eqnarray}
and putting $C=B/2$ -- absence of radiation, cf. Eq. (\ref{bootstrap}), (without radiation in Lorentzian signature spacetime this solution was derived in \cite{Shapiroetal,Shapiroetal1}). This equation then reduces to \cite{hatch}
    \begin{eqnarray}
    \dot{a}^2 =1-H_\pm^2 a^2,\quad
    H_\pm^2=\frac{1\pm\sqrt{1-2BH^2}}B=
    \left.\frac1{a_\mp^2}\,\right|_{\,C=B/2}.\label{efeq3}
    \end{eqnarray}
Obviously the solutions to these two equations, $a(\tau)=\sin(H_\pm\tau)/H_\pm$, represent spherical Euclidean instantons $S_\pm^4$ of the radii $a_\mp$ respectively, or the strings of such spheres touching each other at their poles and forming a ``{\em necklace}" with any number of such spherical beads \cite{hatch}. Note that the value of $C=B/2$ is consistent with the bootstrap equation (\ref{bootstrap}), because the time period for such a necklace consisting of $m$ beads,
    \begin{eqnarray}
    \eta=2m\int_0^{a_\pm}
    \frac{da}{\dot{a}a}=\infty,                       \label{period2}
    \end{eqnarray}
diverges at the poles of spherical beads, where they touch each other -- the range of integration over $a$ in contrast to Eq.(\ref{period1}) is a multiple of the range between $a=0$ at the pole of the 4-sphere $S_\pm^4$ and its value $a_\mp$ at the equator of $S_\pm^4$. Therefore both $F(\eta)$ and $dF(\eta)/d\eta$ vanish and give in view of the bootstrap equation (\ref{bootstrap}) the value of $C=B/2$.

These vacuum (or zero temperature, $1/\eta=0$) necklace instantons existing for all values of $\varLambda=3H^2>0$ are, however, not interesting because their contribution to the statistical sum is suppressed to zero by their infinite {\em positive} Euclidean action. For $B>0$ the on-shell value of the action (\ref{effaction0}),
    \begin{equation}
    \varGamma_0= F(\eta)-\eta F'(\eta)
    +4m_P^2\int_{S^1}
    \frac{d\tau}{a}\,\dot{a}^2\Big(B-a^2
    -\frac{B\dot{a}^2}{3}\Big)\to+\infty,       \label{action-instanton}
    \end{equation}
diverges to $+\infty$ at the poles of necklace beads with $a=0$, where $|\dot a|=1$ and $B-B\dot a^2/3>0$. Thus the CFT cosmology scenario is free from infrared catastrophe of vacuum no-boundary instantons, which would otherwise have a {\em negative} tree-level Euclidean action (proportional to $-1/\varLambda\to -\infty$ at $\varLambda\to 0$) and which would imply that the origin of an infinitely big Universe is infinitely more probable than that of a finite one. Elimination of this infrared catastrophe is the quantum effect of the trace anomaly which flips the sign of the effective action and sends it to $+\infty$ \cite{Bar-Kam-land,hatch,CHS}.

\subsection{Inflation and the hierarchy of Planck and inflation scales}
Inflation stage in this model starts after the ``nucleation" of the system from the gravitational instanton at the $2m$-th turning point of the $m$-folded garland, $\tau_*=2m\int_{a_-}^{a_+}da/\dot a$ (cf. Eq. (\ref{period1})). The Lorentzian time history of the scale factor $a_L(t)$ originates by the analytic continuation of the Euclidean solution $a(\tau)$ to $a_L(t)=a(\tau_*+it)$. This leads to the replacement of oscillatory behavior of $a(\tau)$ by quasi-exponentially growing $a_L(t)$.  When solved with respect to $\dot a^2$ the equation (\ref{efeq0}) can be converterd to the form somewhat diferent from (\ref{mainEq}). In the Lorentzian domain with $a(\tau_*+it)=a_L(t)$ and $\dot a^2(\tau_*+it)=-\dot a_L^2(t)$ this form literally reads as general relativistic Friedmann equation,
    \begin{eqnarray}
    &&\frac{\dot a_L^2}{a_L^2}+\frac1{a_L^2}=
    \frac{\varepsilon}{3M^2_{\rm eff}(\varepsilon)},\quad
    \varepsilon=M_P^2\varLambda+\frac1{2\pi^2 a^4_L}
    \,\sum_\omega\frac{\omega}
    {e^{\omega\eta}\mp 1},               \label{AAA}\\
    &&M^2_{\rm eff}(\varepsilon)=
    \frac{M_P^2}2\left(\,1+\sqrt{1
    -\frac{\beta\,\varepsilon}{12\pi^2M_P^4}}\,\right),   \label{BBB}
    \end{eqnarray}
with the effective Planck mass $M_{\rm eff}(\varepsilon)$ depending on the full matter density $\varepsilon$. This matter density includes together with the contribution of the cosmological constant $\varLambda=3H^2$ also the primordial radiation of the conformal cosmology (but does not include Casimir energy totally screened due to the degravitating effect of conformal anomaly in (\ref{efeq0}) \cite{Bar-Kam-Def}).

The above Euclidean-Lorentzian scenario remains valid also when the cosmological constant is effectively represented by an appropriate potential, $\varLambda\to V(\phi)/M_P^2$, of a slowly varying scalar field $\phi$ playing the role of inflaton \cite{Bar-Kam-land1,Bar-Kam-Nes1}. Remarkably, this scenario automatically leads to the beginning of the inflationary evolution in the vicinity of the {\em maximum} of the potential. Thus it resolves the main difficulty of the no-boundary prescription, according to which under the identification of $\varLambda$ with the value of the potential the most probable is the creation of the Universe at the minimum of $V(\phi)$, see Eq.(\ref{HH}). A vicinity of the maximum of the inflaton potential arises by a very simple mechanism. Obviously, the evolution of the inflaton $\phi$ in the Euclidean domain should also be periodic and subject to the Euclidean equation of motion, which can be rewritten in the form
\begin{equation}
\frac{d}{d\tau}(a^3\dot\phi)=a^3\frac{\partial V}{\partial\phi},
\end{equation}
whence by integrating this equation over the instanton period of oscillations of both $\phi(\tau)$ and $a(\tau)$ one gets 
\begin{equation}
\oint d\tau\,a^3\frac{\partial V}{\partial\phi}=0,
\end{equation}
which means that the $\partial V/\partial\phi$ changes sign at some point inside the instanton domain, so that the instanton is always located at the extremum of the potential. We know that the evolution of both $a(\tau)$ and $\phi(\tau)$ takes place in the Euclidean time, so that $\phi(\tau)$ oscillates between two turning points in the underbarrier regime, which is possible only under the maximum of $V(\phi)$. This is a mechanism of hill-top inflation starting from the nucleation of the system from the Euclidean instanton in the vicinity of the potential maximum.\footnote{Note that this mechanism  resolves the old problem of hill-top inflation, when the inflaton classically rolls directly from the top of the potential and generates infinitely large CMB amplitude. Here after nucleation from the cosmological instanton the inflaton appears on the slope of the potential and, thus, generates a finite amplitude of CMB spectrum \cite{Mukhanov_etal}.}

After the nucleation the evolution consists in the fast quasi-exponential expansion during which the primordial radiation gets diluted, the inflaton field and its energy density $V(\phi)$ slowly decay by a conventional exit scenario and go over into the quanta of conformally non-invariant fields produced from the vacuum.\footnote{\label{creation}  A realistic model should contain a sector of non-conformal fields which can be negligible on top of conformal fields in the early Universe but eventually starts dominating in the course of cosmological expansion.} They get thermalized and reheated to give a new post-inflationary radiation with a sub-Planckian energy density, $\varepsilon\to\varepsilon_{\rm rad}\ll M_P^4/\beta$. Therefore, $M_{\rm eff}$ tends to $M_P$, and one obtains a standard general relativistic inflationary scenario for which initial conditions were prepared by the garland instanton of the above type.

Interestingly, this model can serve as a source of quantum initial conditions for the Starobinsky $R^2$-inflation and Higgs inflation theory \cite{Bar-Kam-Nes,Bar-Kam-Nes1}, in which the effective $H^2$ is generated respectively by the scalaron and Higgs field. In particular, the observable value of the CMB spectral tilt $n_s\simeq 0.965$ in these models can be related to the exponentially high instanton folding number,
$m\simeq\exp(2\pi/\sqrt{3(1-n_s)})\sim 10^8$, whereas the needed inflation scale in these models $H\sim 10^{-6}M_P$ determines the overall parameter $\beta\sim 10^{13}$ \cite{Bar-Kam-Nes,Bar-Kam-Nes1}. Gigantic value of $\beta$ needed to solve the problem of hierarchy between the Planck and inflation scales comprises the most serious difficulty of this scenario. The hope is that this difficulty can be circumvented by means of a hidden sector of numerous conformal fields \cite{Bar-Kam-land1,CHS}.

Obviously the above formalism is capable of inclusion of such fields by replacing sums over conformal field oscillators $\sum_\omega$ with relevant sums $\sum_s\sum_{\omega_s}$ over spins $s$ and their energies $\omega_s$.
However, a high value of $\beta$ cannot be attained by a contribution of low spin conformal fields $\beta=(1/180)\big(\mathcal{N}_0+11 \mathcal{N}_{1/2}+62 \mathcal{N}_{1}\big)$, unless the numbers $\mathcal{N}_s$ of fields of spin $s$ are tremendously high. On the contrary, this bound on $\beta$ can be reached with a relatively low tower of higher spin fields, because a partial contribution of spin $s$ to $\beta$ grows as $s^6$ \cite{Tseytlin}. The solution of hierarchy problem thus becomes a playground of $1/\mathcal{N}$-expansion theory for large number $\mathcal{N}$ of conformal species. However additional problem arises associated with the fact that such theories acquire reduced gravitational cutoff $\varLambda_{\rm grav}\sim M_P/\sqrt{\mathcal N}$ \cite{Dvali-Gabadadze-etal,cutoff,Dvali-Redi,Dvali}, above which effective field theory stops working, and the needed value of inflation scale might exceed this cutoff. Fortunately, due to a peculiar property that the number of polarizations of higher spin conformal particles $\mathcal{N}\sim s^2$ grows with spin much slower than $\beta\sim s^6$, this difficulty is possible to circumvent. If the hidden sector is built of higher spin conformal fields (CHS) \cite{CHS}, then the known gravitational cutoff $\varLambda_{\rm grav}$ turns out to be several orders of magnitude higher than the inflation scale. This justifies the omission of the graviton loop contribution and the use of the above nonperturbative (conformal anomaly based) method.

The further development of this concept has shown that this model of initial conditions suggests many interesting physical predictions. They include potentially observable thermal imprint on primordial CMB spectrum \cite{Bar-Kam-Def,Bar-thermal}, new type of the hill-top inflation \cite{Bar-Kam-Nes,Bar-Kam-Nes1} arising in the synthesis of Higgs inflationary model with a large non-minimal inflaton coupling considered above, etc. Quite interestingly, for the above picture of hierarchy between the Planck and the inflation scales the thermal correction to the spectral parameter of CMB $\Delta n_s^{\rm thermal}\sim -0.001$, depending on the properties of the so-called hill like inflaton potential, might appear in the third decimal order \cite{Bar-thermal,Bar-Kam-Nes} -- the precision anticipated to be reachable in the next generation of CMB observations following Planck. This means that a potential resolution of the hierarchy problem in the CFT scenario via CHS simultaneously would make measurable the thermal contribution to the CMB red tilt. This contribution will be complementary to the most fundamental observational evidence for inflation theory -- red tilt of the primordial CMB spectrum caused by the deviation of the slow roll evolution from the exact de Sitter scenario \cite{S83,Mukhanov_etal}.

Finally, there is one more important comment on this model of initial conditions. Note that this scenario is possible only for spatially closed cosmologies, because it is the presence of the positive curvature term $k/a^2$, $k=+1$, that allows the existence of two turning points for the solution of effective Euclidean equations of motion. For spatially flat or open model with $k=0$ or $-1$ both turning point, do not exist for positive real $a^2$ (note that Eq.(\ref{apm}) then goes over into $a_\pm^2=(k\pm\sqrt{1-4CH^2})/2H^2<0$), so that there is no transition from the periodic motion in classically forbidden domain to the phase of cosmological expansion with the Lorentzian spacetime signature. Phenomenologically this sounds disturbing because inflation is usually assumed to be considered in spatially flat Universe, and and its flatness is considered as one of the advantages of the inflation scenario, matching very well with observations. Thus, the idea of non-flat inflationary cosmology is not popular, even though relevant models of inflation have been worked out \cite{Linde1995,Tanaka-Sasaki1992,Garriga_etal1999,Linde2003}. However, as it is recently observed in the exhaustive treatment of the Planck 2018 CMB temperature and polarization data \cite{Silk_etal2019,Silk_etal2022}, these datasets are now preferring a positive curvature at more than the 99\% confidence level with a mean $\varOmega_K\simeq-0.04$. Even though this preference of closed Universe is associated with discordances arising for local cosmological observables, known as Hubble tension \cite{Hubble_tension}, robust observational evidence in favor of positive spatial curvature serves a strong motivation for the suggested model of quantum initial conditions.

\section{Concluding remarks}

We have presented various aspects and applications of perturbative quantum gravity in the physics of early Universe and quantum cosmology. Starting from somewhat naive normalization requirements for the cosmological wave function, establishing certain restrictions on particle phenomenology, we have amounted to the links between physics of cosmic microwave background at hundreds of megaparsecs scale and physics of electroweak sector of Standard Model of particles and fields at $10^{-16}$ cm scale. The established relation between the parameters of CMB spectra and Higgs boson mass clearly violates a well-known physics principle of separation of scales and exists entirely due to such a perfect microscope existing in Nature as cosmological expansion -- apparently the most significant phenomenon in the Universe.

Moving further down to the past in the history of Universe we put forward a conjecture of its microcanonical density matrix as a quantum state which describes quantum origin of inflationary stage of cosmological evolution. Its construction, conceptually stemming from the Occam razor principle of avoiding redundant assumptions, is based on the path integral representation of a generic solution of Wheeler-DeWitt equations and turns out to be very productive in application to the Universe dominated by conformally coupled matter fields. The density matrix paradigm allows one to solve such a problem as elimination of infrared catastrophe in the theory of the no-boundary quantum state of the Universe and suggests initial conditions for inflation in the form of a special quasi-thermal cosmological instanton.

Conceptually the microcanonical density matrix of the Universe suggests the reincarnation of the Big Bang scenario on a qualitatively new level of its understanding. Inflation paradigm replaced the notion of infinitely hot and dense initial state of the Universe by an effectively vacuum state at zero temperature, whose quantum fluctuations eventually gave rise to large scale structure of the Universe. In the density matrix paradigm we again get back to the notion of initially hot Universe originating from the classically forbidden state of the gravitational field, but characterized by finite rationally manageable parameters of effective temperature, size and rate of expansion.

Quantum origin of the Universe might be sub-Planckian, even relatively low energy, phenomenon, whose description then may be possible within our semiclassical perturbative methods. The success of this conjecture depends on the construction of a hypothetical particle physics model of numerous higher spin conformal fields, which would form a theoretical framework for the solution of quantum initial conditions problem in cosmology. Moreover, as it stands it also might serve as a selection criterion for the landscape of stringy vacua. String theory, as is well known, incorporates an infinite tower of higher spin fields which, however, are massive and cannot be conformally coupled to spacetime metric. On the other hand, there is a popular idea that strings might be a broken phase of the higher spin theory \cite{Sagnotti, Vasiliev}. This might open prospects of unification of the suggested ideas with the fundamental origin of particle physics phenomenology necessary for the solution of Planck vs inflation scale hierarchy problem.

\section*{Acknowledgement}
We are grateful to E.~Alvarez, J.~Barbon, F. Bezrukov, I. L. Buchbinder, R. Casadio, C. Deffayet, D. Diakonov, G. Dvali, G. Esposito, V. P. Frolov, J.~Garriga, G.~Gibbons, I.~Ginzburg, J. B. Hartle, N.~Kaloper,  I. P. Karmazin, I.~M.~Khalatnikov, C. Kiefer, I. V. Mishakov, E. Mottola, V. Mukhanov, D. V. Nesterov, D. Page, V. N. Ponomariov, V. A. Rubakov, I. Sachs, I. L. Shapiro, M. Shaposhnikov, D.~V.~Shirkov, S.~Solodukhin, P. C. E. Stamp, A. A. Starobinsky, C. F. Steinwachs, O. V. Teryaev, A. Tronconi, A. A. Tseytlin, H. Tye, G.~P.~Vacca, T. Vardanyan, M. A. Vasiliev, G.~Venturi, A. Vilenkin, G. A. Vilkovisky, W. Unruh and R.~Woodard for fruitful discussions. The work of A.B. was partially supported by the Russian Foundation for Basic Research grant 20-02-00297 and by the Foundation for Theoretical Physics Development ``Basis''.

%
%
%



\end{document}